\documentclass[12pt]{article}
\usepackage[utf8]{inputenc}
\usepackage{amsmath,amsfonts,amssymb,epsfig,fp,color,bbm,tikz-cd,url,hyperref}
\def\be{\begin{equation}\begin{gathered}}
\def\ee{\end{gathered}\end{equation}}
\def\Sum{\sum\limits}
\def\Prod{\prod\limits}
\def\half{{\textstyle{\frac12}}}
\def\d{\partial}
\def\pd{\partial}

\def\tr{\mathrm{\,tr\,}}

\def\rank{{\rm rank}}
\def\bs{\boldsymbol}
\def\mf{\mathfrak}
\def\mc{\mathcal}
\def\mr{\mathrm}
\def\mb{\mathbb}
\def\Oint{\oint\limits}
\def\Res{\mathrm{\,Res\,}}
\def\stackreb#1#2{\mathrel{\mathop{#2}\limits_{#1}}}
\newcommand{\rf}[1]{(\ref{#1})}

\newcommand{\eq}[1]{\begin{equation}\begin{gathered}
#1\end{gathered}\end{equation}}
\newcommand{\eqs}[1]{\begin{equation}\begin{gathered}\begin{split}
#1\end{split}\end{gathered}\end{equation}}
\newcommand{\bpm}{\begin{pmatrix}}
\newcommand{\epm}{\end{pmatrix}}

\newtheorem{theorem}{Theorem}
\newtheorem{lemma}{Lemma}
\newtheorem{definition}{Definition}

\numberwithin{equation}{section}

\begin{document}

\title{Twist-field representations of W-algebras, exact conformal blocks and character identities}

\author{M.~Bershtein, P.~Gavrylenko, A.~Marshakov}

\date{}

\maketitle

\begin{abstract}
\noindent
We study the twist-field representations of W-algebras and generalize construction of the corresponding vertex operators to D- and B-series. It is shown, how the computation of characters of these representations leads to nontrivial identities involving lattice theta-functions. We also propose a way to calculate their exact conformal blocks, expressing them for D-series in terms of geometric data of the corresponding Prym
variety for covering curve with involution.

\end{abstract}

\tableofcontents

\section{Introduction}

Representation theory of W-algebras \cite{ZFLW} is still a subject with many open questions and even challenges. These questions often arise in the context of two-dimensional conformal field theory (CFT) with extended symmetry, and due to recently found nontrivial correspondence \cite{LMN,NO,AGT,BPSCFT} can be also important even for multidimensional supersymmetric gauge theories~\footnote{In addition to supersymmetric gauge theories there are other important multidimensional applications of W-algebras, including at least the holographic dualities, AdS-type geometries and Chern-Simons gauge theories, see e.g. \cite{R1,R2,R3} and references therein. However, the role of particular twist-field representations of W-algebras, considered below -- especially for other Dynkin series, for these applications is much less clear, than their possible role for supersymmetric gauge theories.}.

The main object of this study is a conformal block. Generally, the space of conformal blocks is the space of all functionals on the product of Verma modules at marked points on the Riemann surface, that solve the Ward identities (see e.g. \cite{Frenkel BenZvi}, where the language of coinvariants was used). Such spaces form certain bundles over the moduli space of complex curves with marked points.

However, in order to construct the correlation functions in physical theories one should be able to extract a particular section of this bundle.
The corresponding functionals are computed usually on the highest weight vectors in each of the Verma modules. For the Virasoro algebra the conformal blocks are specified by particular intermediate dimensions, or equivalently, by the asymptotic behavior of a
conformal block on the boundary of moduli space \cite{BPZ}. It turns out that for generic W-algebra it is not enough to fix quantum numbers in the intermediate channels. Even for sphere with three marked points the vector space of conformal blocks becomes infinite dimensional for $W_N$ algebras with $N>2$.

However, for certain particular cases this conformal block can be constructed explicitly applying some extra machinery. In what follows we restrict ourselves to the case of integer (and sometimes -- half-integer) Virasoro central charges, when representations of W-algebras are more directly related to the representations of the corresponding Kac-Moody (KM) algebras (of level $\texttt{k}=1$), and the corresponding quantum field theories can be directly described by (minimal possible amount of) free fields \cite{FK}.

One of the recent methods \cite{GMfer} reduces
the problem here to the Riemann-Hilbert problem, arising in the context of the isomonodromy/CFT correspondence \cite{GIL,ILT,GIso}. The
key idea of this approach is to extract the concrete conformal block by implying the condition that it has predictable monodromies after insertion of the
simplest degenerate fields, or fermions. Morally it means, that we fix the 3-point blocks as eigenvectors of the Verlinde loop operators and parameterize them by the monodromy data.

Even in such situation, in case of generic monodromies one cannot write explicit formula for the conformal block. Below, following \cite{GMtw}, we are going to restrict ourselves to the case of so-called twist fields \cite{ZamAT},
corresponding to the quasi-permutation monodromies, when the representations of W-algebras become related to the twisted
representations of the corresponding KM algebras~\footnote{To prevent terminological confusion we should stress that ``twist field representation'' is different from
``twisted representation'': the latter implies that algebra is changed (twisted!) itself, whereas the first only reflects the way --- how this
representation was constructed.} \cite{KacBook,Bakalov:2004}.

The paper is organized as follows. We start from formulation of the representations of KM and W-algebras in terms of free bosonic and fermionic fields. First the $GL(N)$ case is revisited, and we extend it to the $D$- and $B$- series, using the real fermions. Then we define the twist field representations, corresponding to the elements of Cartan's normalizer $g\in\mr N_{G}(\mf h)$ together with some extra data (to be denoted as $\dot{g}$ and $\ddot{g}$, see the details in Sect.~\ref{ss:twists} below). Bosonization of the twist fields essentially depends on the conjugacy classes in $\mr N_{G}(\mf h)$. We classify therefore these conjugacy classes for  $G=GL(N)$ and $G=O(n)$ (for $n=2N$ and $n=2N+1$) and define the twist fields $\mathcal{O}_{\ddot g}$ in
terms of the boundary conditions 
in corresponding free field theory.

Bosonization rules allow to compute straightforwardly the characters $\chi_{\dot{g}}(q)$ of the corresponding representations. For the twist fields of ``$GL(N)$ type'' this goes back to the seminal results of Al.~Zamolodchikov and V.~Knizhnik, and we develop here a similar technique in the case of real fermions and another class of the twist fields, arising only for $D$- and $B$- series. The character formulas include summations over the root lattices, reflecting the fact that one deals here with the class of lattice vertex algebras. Dependently on the conjugacy class $g\in\mr N_{G}(\mf h)$ of a twist field, the lattice can be reduced to its projection to the Weyl-invariant part: in such cases ``smaller'' lattice theta functions show up, or we find even a kind of ``duality'' between the lattice theta functions for $D$- and $B$- series.

If two
different elements $g_{1,2}\in\mr N_{G}(\mf h)$ are nevertheless conjugated $g_1\sim g_2$ in $G$, but not
in $\mr N_{G}(\mf h)$, this gives (for appropriate additional data) a nontrivial identity $\chi_{\dot{g}_1}(q) = \chi_{\dot{g}_2}(q)$ between two characters, involving lattice theta-functions. Such character identities go back to 70's of the last century (see \cite{Macdonald:1972}, \cite{KacId}) and even to Gauss, but our derivation gives probably new identities, involving in particular the theta functions for $D$- and $B$-root lattices.

Finally, we propose a construction of the exact conformal blocks of the twist fields for W-algebras of $D$-series, generalizing approach of \cite{ZamAT,GMtw}, and obtain explicit formula, expressing the
multipoint blocks in terms of algebro-geometric objects on the branched cover of the original curve with extra involution. Some important technical details are collected in Appendices.

\section{W-algebras and KM algebras at level one
\label{ss:WandKM}}

\subsection{Boson-fermion construction for GL(N)}

We start from the standard complex fermions
\eq{
\psi^*_\alpha(z)=\Sum_{p\in\frac12+\mathbb Z}\frac{\psi^*_{\alpha,p}}{z^{p+\frac12}},\ \ \ \
\psi_\alpha(z)=\Sum_{p\in\frac12+\mathbb Z}\frac{\psi_{\alpha,p}}{z^{p+\frac12}}
}
with the operator product expansions (OPE's)
\be
\psi^*_\alpha(z)\psi_\beta(w)=-\psi_\beta(w)\psi^*_\alpha(z)=\frac{\delta_{\alpha\beta}}{z-w}+reg.\\
\psi_\alpha(z)\psi_\beta(w)=\psi^*_\alpha(z)\psi^*_\beta(w)=reg.\label{OPExi}
\ee
equivalent to the following anticommutation relations
\eq{
\{\psi^*_{\alpha,p},\psi_{\beta,q}\}=\delta_{\alpha\beta}\delta_{p+q,0},\quad \{\psi_{\alpha,p},\psi_{\beta,q}\}=\{\psi^*_{\alpha,p},\psi^*_{\beta,q}\}=0,\ \ \ \ p,q\in\half+\mathbb Z
}
One can generate the Kac-Moody $\widehat{\mathfrak{gl}(N)}_1$ algebra by the currents
\eq{
J_{\alpha\beta}(z)=:\psi_{\alpha}^*(z)\psi_\beta(z):
}
where the free fermion normal ordering moves all $\{\psi_r\}$ and $\{\psi^*_r\}$ with $r>0$ to the right.
These currents have standard OPE's:
\eq{
J_{\alpha\beta}(z)J_{\gamma\delta}(w)=\frac{\delta_{\beta\gamma}\delta_{\alpha\delta}}{(z-w)^2}+\frac{\delta_{\beta\gamma}J_{\alpha\delta}(w)-
\delta_{\alpha\delta}J_{\beta\gamma}(w)}{z-w}+reg.
}
and when expanded into the (integer!) powers of $z$
\eq{
J_{\alpha\beta}(z)=\Sum_{n\in\mathbb Z}\frac{J_{\alpha\beta,n}}{z^{n+1}}
}
one gets the following Lie algebra commutation relations
\eq{
\left[J_{\alpha\beta,n},J_{\gamma\delta,m}\right]=n\delta_{n+m,0}\delta_{\beta\gamma}\delta_{\alpha\delta}+\delta_{\beta\gamma}J_{\alpha\delta,m+n}-
\delta_{\alpha\delta}J_{\gamma\beta,m+n},\ \ \ \ n,m\in \mathbb{Z}
\label{commJ}
}
This set of generators contains the zero modes $J_{\alpha\beta,0}$, generating the $\mathfrak{gl}(N)\subset\widehat{\mathfrak{gl}(N)}_1$ subalgebra. The $W(\mf{gl}(N))= W_N\oplus \sf H$ 
algebra~\footnote{Here $\sf H=\widehat{\mathfrak{gl}(1)}_1$ is the Heisenberg algebra, and $W_N=W(\mf{sl}(N))$.}
can be defined in a well-known way -- as a commutant of $\mathfrak{gl}(N)$ in the (completion of the)
universal enveloping algebra $U(\widehat{\mathfrak{gl}(N)}_1)$.

There are several different sets of generators of the $W(\mf{gl}(N))= W_N\oplus \sf H$ algebra. In what follows mostly convenient for our purposes is to use the fermionic bilinears
\be
\Sum_{\alpha=1}^N \psi^*_\alpha\left(z+\half t\right)\psi_\alpha\left(z-\half t\right)=\frac Nt+\Sum_{k=1}^\infty\frac{t^{k-1}}{(k-1)!}
U_k(z)
\label{Wdef}
\ee
or, in terms of the Hirota derivative $D_z^n f(z)\cdot g(z) = \left.\left(\d_{z_1}-\d_{z_2}\right)^nf(z_1)g(z_2)\right|_{z_1=z_2=z}$,
\eq{
U_k(z)=D_z^{k-1}\Sum_{\alpha=1}^N :\psi^*_\alpha(z)\cdot\psi_\alpha(z):
\label{Wgln}
}
Another useful set of generators is the bosonic representation
\eq{\label{eq:Wk:Ja}
W_k(z)=\Sum_{\alpha_1<\ldots <\alpha_k}:J_{\alpha_1\alpha_1}(z)\ldots J_{\alpha_k\alpha_k}(z):
\equiv \Sum_{\alpha_1<\ldots <\alpha_k}:J_{\alpha_1}(z)\ldots J_{\alpha_k}(z):
}
equivalent to the quantum Miura transform from \cite{ZFLW}. To explain that \eqref{Wgln} is actually
equivalent to \eqref{eq:Wk:Ja} one can use already noticed description of $W(\mf{gl}(N))$ as the centralizer of the screening operators, which generate
$\mathfrak{gl}(N)\subset\widehat{\mathfrak{gl}(N)}_1$. Since it is already known that \eqref{eq:Wk:Ja} is the centralizer of the screening operators \cite{FF}, it remains to show the same for
\eqref{Wgln}, what can be done in several steps:
\begin{enumerate}
\item Consider all normally-ordered fermionic monomials $:\prod_i \d^{k_i}\psi_{\alpha_i}(z) \prod_i \d^{l_i}\psi^*_{\beta_i}(z):$, which transform
as tensors under the action of $GL(N)$. By \emph{the first fundamental theorem} of invariant theory \cite{Weyl} the only invariants in such representation
are given by all possible contractions, so they can be written as
$:\prod_i(\Sum_\alpha\d^{k_i}\psi_{\alpha}(z)\d^{l_i}\psi^*_{\alpha}(z)):$.
\item Any such expression can be obtained by taking the regular products (constant terms in the OPE, denoted by brackets) of the ``elementary elements''
 $:\Sum_\alpha\d^{k}\psi_{\alpha}(z)\d^{l}\psi^*_{\alpha}(z):$, since
\be
\left(:\prod_i\Sum_\alpha\d^{k_i}\psi_{\alpha}(z)\d^{l_i}\psi^*_{\alpha}(z):\, :\Sum_\beta\d^{k}\psi_{\beta}(z)\d^{l}\psi^*_{\beta}(z):\right)=\\=
:\prod_i\Sum_\alpha\d^{k_i}\psi_{\alpha}(z)\d^{l_i}\psi^*_{\alpha}(z)
\Sum_\beta\d^{k}\psi_{\beta}(z)\d^{l}\psi^*_{\beta}(z):+\text{ lower terms in $\psi,\psi^*$}
\ee
Therefore one can perform this procedure iteratively and express everything as regular products of bilinears.
\item Any element $:\Sum_\alpha\d^{k}\psi_{\alpha}(z)\d^{l}\psi^*_{\alpha}(z):$ can be expressed as a linear combination of
$\d^{l'+1}U_{k'}(z)$ for different $l'$ and $k'$ with $l'+k'=l+k$.

\end{enumerate}

Hence, the generators $\{U_k(z)\}$ are expressible in terms of $\{W_k(z)\}$ (and vice versa) by some non-linear triangular
transformations, but we do not need here its explicit form~\footnote{
The fact that nonlinear W-algebra generators can be expressed through just bilinear fermionic expressions is well-known, and was already exploited in \cite{LMN,NO} (see also \cite{GMfer} and references therein).}.

Formally there is an infinite number of generators in \rf{Wdef} and \rf{Wgln}, but since all of them are expressed in terms of $N$ generators
\eqref{eq:Wk:Ja}, they are not independent: one always has
\be
U_{N+n}(z)=P_n(\{\d^k U_{l\leq N}\})
\ee
for some polynomials $\{P_n\}$, and this is the origin of non-linearity of 
W-algebra \cite{FKRW}. The relation between fermions and bosons is given by well-known \cite{FK,KVdL} bosonization formulas
\be
\psi^*_\alpha(z)=\exp\left(-\Sum_{n<0}\frac{J_{\alpha,n}}{nz^{n}}\right)\exp
\left(-\Sum_{n>0}\frac{J_{\alpha,n}}{nz^{n}}\right)e^{Q_\alpha} z^{J_{\alpha,0}}\epsilon_\alpha(\bs J_0)=
\\
= e^{i\varphi_{\alpha,-}(z)}e^{i\varphi_{\alpha,+}(z)}e^{Q_\alpha}z^{J_{\alpha,0}}\epsilon_\alpha(\bs J_0)
\\
\psi_\alpha(z)=\exp\left(\Sum_{n<0}\frac{J_{\alpha,n}}{nz^{n}}\right)\exp
\left(\Sum_{n>0}\frac{J_{\alpha,n}}{nz^{n}}\right)e^{-Q_\alpha} z^{-J_{\alpha,0}}\epsilon_\alpha(\bs J_0)=
\\
= e^{-i\varphi_{\alpha,-}(z)}e^{-i\varphi_{\alpha,+}(z)}e^{-Q_\alpha}z^{-J_{\alpha,0}}\epsilon_\alpha(\bs J_0)
\label{bosonization}
\ee
where $\epsilon_\alpha(\bs J_0)=\prod_{\beta=1}^{\alpha-1}(-1)^{J_{0,\beta}}$, and diagonal $J_{\alpha\alpha,n} \equiv J_{\alpha,n}$ form the Heisenberg algebra:
\eq{
\left[J_{\alpha,n},J_{\beta,m}\right]=n\delta_{\alpha\beta}\delta_{m+n,0},
\ \ \ \ \ \ \ [J_{\alpha,0},Q_\beta]=\delta_{\alpha\beta}
\label{heisenberg}
}
One can also express all other generators in terms of (positive and negative parts of) the bosonic fields
\eq{
i\varphi_{+,\alpha}(z)=-\Sum_{n>0}\frac{J_{\alpha,n}}{n z^n},\ \ \ \
i\varphi_{-,\alpha}(z)=-\Sum_{n<0}\frac{J_{\alpha,n}}{n z^n}
}
namely
\be J_{\alpha\beta}(z)=e^{i\varphi_{-,\alpha}-i\varphi_{-,\beta}}e^{i\varphi_{+,\alpha}-i\varphi_{+,\beta}}e^{Q_\alpha-Q_\beta}z^{J_{\alpha,0}
-J_{\beta,0}}(-1)^{
\Sum_{\gamma=\alpha-1}^{\beta-1}J_{\gamma,0}+\theta(\beta-\alpha)},\ \ \ \alpha\neq\beta \\
J_{\alpha\alpha}(z)=J_\alpha(z) = \frac{J_{\alpha,0}}z+ i\d\varphi_{+,\alpha}(z) + i\d\varphi_{-,\alpha}(z)
\label{Jall}
\ee

\subsection{Real fermions for $D$- and $B$- series}

Let us now almost repeat the same construction for the orthogonal series $B_N$ and $D_N$, which correspond to  W-algebras $W(\mf{so}(2N+1))$ and $W(\mf{so}(2N))$ respectively.
The corresponding KM algebras at level one can be realized in terms of the real fermions (see e.g. \cite{Windey}) with the OPE's
\eq{
\Psi_i(z)\Psi_j(w)=\frac{\delta_{ij}}{z-w}+reg.,\ \ \ \ \ i,j=1,\ldots,n
\label{reope}
}
(here dependently on the case we put either $n=2N$ or $n=2N+1$),
which corresponds to anti-commutation relations
\eq{
\{\Psi_{i,p},\Psi_{j,q}\}=\delta_{ij}\delta_{p+q,0},\ \ \ \ \ \ p,q\in\half+\mathbb Z
}
One can say that these OPE's and commutation relations are defined by metric 
on $n$-dimensional Euclidean space, or
symbolically by $ds^2=\Sum_{i=1}^n d\Psi_i^2$. The KM currents are again expressed by bilinear combinations
\eq{
J^{(1)}_{ik}(z)=:\Psi_i(z)\Psi_j(z):
\label{Iso}
}
and satisfy usual commutation relations together with $J^{(1)}_{ij}(z)=-J^{(1)}_{ji}(z)$. It is also convenient to pass to the complexified fermions ($\alpha=1,\ldots,N$)
\eq{
\psi^*_{\alpha}(z)=\frac1{\sqrt2}\left(\Psi_{2\alpha-1}(z)+i\Psi_{2\alpha}(z)\right),
\ \ \ \ \
\psi_{\alpha}(z)=\frac1{\sqrt2}\left(\Psi_{2\alpha-1}(z)-i\Psi_{2\alpha}(z)\right)
\label{compexif}
}
which due to \rf{reope} have the standard OPE's given by \rf{OPExi}. Let us point out that $B_N$-series ($i,j=1,\ldots,2N+1$) differs from $D_N$-series ($i,j=1,\ldots,2N$) by remaining single real fermion $\Psi_{2N+1}(z)=\Psi(z)$.

Using the complexified fermions the generators \rf{Iso} can be re-written as
\be
J_{\alpha\beta}=-J_{\bar\beta\bar\alpha}=:\psi_\alpha^*(z)\psi_\beta(z): = \frac12(J^{(1)}_{2\alpha-1,2\beta-1}+J^{(1)}_{2\alpha,2\beta})
+ \frac{i}2(J^{(1)}_{2\alpha,2\beta-1}+J^{(1)}_{2\beta,2\alpha-1})
\ee
together with
\be
J_{\alpha\bar\beta}=\psi_\alpha^*(z)\psi_\beta^*(z),\ \ \
J_{\bar\alpha\beta}=\psi_\alpha(z)\psi_\beta(z)\\
J_{\alpha,\Psi}=\psi^*_\alpha(z)\Psi(z),\ \ \
J_{\bar\alpha,\Psi}=\psi_\alpha(z)\Psi(z)
\ee
so that we see explicitly $\widehat{\mf{gl}(N)}_1\subset\widehat{\mf{so}(n)}_1$. Note also that elements $J_{\alpha\alpha}(z)=J_\alpha(z)$ again form the Heisenberg algebra, and its
zero modes $J_{\alpha,0}$ correspond to the Cartan subalgebra of $\mf{so}(n)$.

As before, we define the W-algebra $W(\mf{so}(n))$ as the commutant of $\mf{so}(n)\subset\widehat{\mf{so}(n)}_1$. In contrast to the simply-laced cases,
one finds this commutant for $B$-series being not in the completion of $U\left(\widehat{\mf{so}(2N+1)}_1\right)$, but rather in the entire fermionic algebra.
 An inclusion of algebras $\mf{gl}(N)\subset\mf{so}(2N)$ acting on the
same space leads to inverse inclusion
\eq{
W(\mf{so}(2N))\subset W(\mf{gl}(N))
}
Similarly to \rf{Wgln} one can present the generators of the $W(\mf{so}(n))$-algebra explicitly, using the real fermions
\eq{\label{eq:WgenBD}
U_k(z)=\frac12 D_z^{k-1}\Sum_{j=1}^n:\Psi_j(z)\cdot\Psi_j(z):,\ \ \ \ \
V(z)=\prod_{j=1}^n\Psi_j(z)
}
The last current is bosonic in the $D_N$ case, but fermionic for the $B_N$-series. These expressions are obtained analogously to \eqref{Wgln} with the help of
invariant theory, the only important difference is that for $SO(n)$ case there is also completely antisymmetric invariant tensor.
We can rewrite these expressions using complex fermions (for the $D_N$ case one should just put here $\Psi(z)=0$ in the expressions for $U$-currents and $\Psi(z)=1$ in the expressions for the $V$-current)
\eq{
U_k(z)=\half D_z^{k-1}\Sum_{\alpha=1}^N\left(\psi^*_\alpha(z)\cdot\psi_\alpha(z)+
\psi_\alpha(z)\cdot\psi_\alpha^*(z)\right)+\half D_z^{k-1}\Psi(z)\cdot\Psi(z)\\
V(z)=\Prod_{\alpha=1}^N:\psi_\alpha^*(z)\psi_\alpha(z):\Psi(z)
}
It is easy to see that odd generators vanish $U_{2k+1}=0$, while even generators in the $D_N$ case coincide with those in $W(\mf{gl}(N))$ algebra
\eq{
U_{2k}(z)=D_z^{2k-1}\Sum_{\alpha=1}^{N}:\psi_\alpha^*(z)\cdot\psi_\alpha(z):+
\half D_z^{2k-1}\Psi(z)\cdot\Psi(z),\ \ \ k=1,2,\ldots
\label{Wson}
}
Hence, finally we have the following sets of independent generators:
\eqs{
U_1(z),U_2(z),\ldots, U_N(z)&\quad \text{for } W(\mf{gl}(N))\\
U_2(z),U_4(z),\ldots, U_{2N-2}(z), V(z)&\quad \text{for } W(\mf{so}(2N))\\
U_2(z),U_4(z),\ldots, U_{2N}(z), V(z)&\quad  \text{for } W(\mf{so}(2N+1))
\label{currents}
}

\section{Twist-field representations from twisted fermions
\label{ss:twists}}
\subsection{Twisted representations and twist-fields}
\label{ssec:twistmain}

For any current algebra, generated by currents $\{ \Phi_I(z)\}$, the commutation
relations follow from their local OPE's
\eq{
\Phi_I(z)\Phi_J(w)\ \stackreb{z\to w}{=}\ \Sum_{K}\frac{(\Phi_I\Phi_J)_K(w)}{(z-w)^K}
}
However, to define the commutation relations, in addition to local OPE's one should also know the boundary
conditions for the currents: in radial quantization --- the analytic behaviour of $\Phi_I(z)$ around zero.
Any vertex operator $\mc V_g(0)$, e.g. sitting at the origin~\footnote{For nontrivial boundary conditions we assume the presence of such field by default, when obvious --- not indicating it explicitly.}, can create a
nontrivial monodromy for our currents:
\eq{
\Phi_I(e^{2\pi i}z)\mc V_g(0)=\Sum_jg_{IJ}\Phi_J(z)\mc V_g(0)
}
or some linear automorphism of the current algebra.

\paragraph{Example:} Perhaps the simplest example of such nontrivial monodromy is the diagonal transformation of 
fermionic fields
\be
\psi^*_\alpha(e^{2\pi i}z) = e^{2\pi i\theta_\alpha}\psi^*_\alpha(z),\ \ \
\psi_\alpha(e^{2\pi i}z) = e^{-2\pi i\theta_\alpha}\psi_\alpha(z),\ \ \ \alpha=1,\ldots,N
\label{diagmn}
\ee
which comes just from the shifts the mode expansion indices
\eq{
\psi^*_\alpha(z)=\Sum_{p\in\mathbb Z+\frac12}\frac{\psi^*_{\alpha,p}}{z^{p+\frac12-\theta_\alpha}},\ \ \ \
\psi_\alpha(z)=\Sum_{p\in\mathbb Z+\frac12}\frac{\psi^*_{\alpha,p}}{z^{p+\frac12+\theta_\alpha}}
\label{fers}
}
Instead of \rf{OPExi} one gets therefore
\eq{
\psi^*_\alpha(z)\psi_\beta(w) \rightarrow
z^{\theta_\alpha}w^{-\theta_\beta}\psi^*_\alpha(z)\psi_\beta(w)=
\frac{\delta_{\alpha\beta}z^{\theta_\alpha}w^{-\theta_\beta}}{z-w}+z^{\theta_\alpha}w^{-\theta_\beta}
:\psi^*_\alpha(z)\psi_\beta(w):\ =
\\
\stackreb{z\to w}{=}\ \frac{\delta_{\alpha\beta}}{z-w}+\frac{\theta_\alpha\delta_{\alpha\beta}}{w}+:\psi^*_\alpha(w)\psi_\beta(w):+reg.
}
which means that for the shifted fermions \rf{fers} one should use the different normal ordering:
\be
\big(\psi^*_\alpha(z)\psi_\beta(z)\big)=\frac{\theta_\alpha\delta_{\alpha\beta}}z+:\psi^*_\alpha(z)\psi_\beta(z):
\ee
This implies that for the diagonal components of the $\widehat{\mf{gl}(N)}_1$ algebra one has extra shift $J_\alpha(z)\rightarrow J_\alpha(z) + \frac{\theta_\alpha}{z}$, while for the non-diagonal currents we
obtain
\eq{
J_{\alpha\beta}(z)=\Sum_{n\in\mathbb Z}\frac{J_{\alpha\beta,n}}{z^{n+1+\theta_\alpha-\theta_\beta}}
}
so that the
commutation relations for this ``twisted'' Kac-Moody algebra become
\eq{
\left[J_{\alpha\beta,n},J_{\gamma\delta,m}\right]=(n-\theta_\alpha+\theta_\beta)
\delta_{n+m,0}\delta_{\beta\gamma}\delta_{\alpha\delta}+\delta_{\beta\gamma}J_{\alpha\delta,m+n}-
\delta_{\alpha\delta}J_{\beta\gamma,m+n}
}
We see that these commutation relations differ from the conventional ones \rf{commJ} only by extra shift, which can be hidden into the Cartan
generators $J_{\alpha\alpha,0}$. However, in the twisted case one cannot think about W-algebra
as the commutant of certain $\mf{gl}(N)$, since there is no such zero mode subalgebra in twisted $\widehat{\mf{gl}(N)}_1$. Nevertheless we define the currents
\eq{
U_k(z)=D_z^{k-1}\Sum_{\alpha=1}^N\big(\psi^*_\alpha(z)\cdot\psi_\alpha(z)\big)
}
and one can still use two basic facts:
\begin{itemize}
  \item since $U_k(e^{2\pi i}z)=U_k(z)$, they are expanded in integer powers of $z$ as before;
  \item they satisfy the same algebraic relations for all values of monodromies $\{\theta_\alpha\}$, because the OPE's of $\psi_\alpha, \psi^*_\alpha$, and thus
the OPE's of $U_k$,
do not depend on these monodromy parameters.
\end{itemize}

Consider now more general situation, when
\eq{\label{eq:gentwist}
\psi^*_\alpha(e^{2\pi i}z) = \Sum_{\beta=1}^N g_{\alpha\beta}\psi^*_\beta(z),\ \ \ \
\psi_\alpha(e^{2\pi i}z) =  \Sum_{\beta=1}^N g^{-1}_{\beta\alpha}\psi_\beta(z)
}
i.e. unlike \rf{diagmn}, the monodromy is no longer diagonal~\footnote{The element $g$ should preserve
the structure of the OPEs, so it should preserve symmetric form on fermions, and therefore lies in $O(2N)$. It automatically implies that all
even generators of the W-algebra $U_{2k}(w)$ are also preserved. To preserve odd generators $U_{2k+1}(z)$ one should also have $g\in Sp(2N)$, but
$O(2N)\cap Sp(2N)=GL(N)$, therefore $g\in GL(N)$.}.
It is clear that then the action on $\widehat{\mf{gl}(N)}_1$ is
\eq{\label{eq:gentwistJ}
J_{\alpha\beta}(z)\mapsto g_{\alpha\alpha'}g^{-1}_{\beta'\beta}J_{\alpha'\beta'}(z)
}
The most general transformation we consider in the $O(n)$ case mixes $\psi$ and $\psi^*$:
\eq{
\psi_\alpha(e^{2\pi i}z) = \Sum_{\beta=-N}^N g_{\alpha\beta}\psi_\beta(z),\ \ \ \ \ \alpha=-N,\ldots,N
}
where it is convenient to introduce conventions $\psi^*_{-\alpha}=\psi_\alpha$, $\alpha>0$, and $\psi_0$ can be absent. Here matrix $g$ should preserve the anticommutation relations.

Consider now a sequence of (super) vertex algebras
\[\text{W-algebra} \;\; \subset \;\; \text{Heisenberg algebra } \widehat{\mathfrak{h}} \;\; \subset \;\; 	 \text{KM-algebra} \;\; \subset \;\; \text{Fermions}.\]
Taking an element $g \in G$ we construct a twisted representation of the fermionic algebra $\psi,\psi^*$. Then, for $G=GL(N)$ or $G=SO(2N)$ one gets the KM algebra as the invariant of the group $\Gamma$, acting on the fermionic algebra ($\Gamma=U(1)$ and $\Gamma=\mathbb{Z}/2\mathbb{Z}$ in two cases correspondingly). This group acts on the twisted representations of the fermionic algebra, therefore the twisted representations of KM algebra are labeled by pairs ($g$, $\Gamma$ isotypic component), in what follows we denote such pair by $\dot{g}$~\footnote{Since the action of $\Gamma$ on the twisted representations of fermionic algebra is not canonical here is no canonical identification of the set $\Gamma$ isotypic components and the set of representations of $\Gamma$. In our case the group $\Gamma$ is commutative, and the set of $\Gamma$ isotypic components forms a principal homogeneous space for group $\operatorname{Rep}\Gamma$, (see also Sect.~3b of \cite{DVVV} for generic discussion of non-uniqueness of the action of $\Gamma$ in orbifold models). However, the definition of the action of $\Gamma$ in our examples will be always clear below.}.
Denote by $\mathcal{H}_{\dot{g}}$ the corresponding representation of the KM algebra for $G=GL(N)$ or $G=SO(2N)$. For the $G=SO(2N+1)$ case $\mathcal{H}_{\dot{g}}$ is a representation of the fermionic algebra itself. An explicit description of $\mathcal{H}_{\dot{g}}$ and calculation of its characters is given in Sect.~\ref{sec:characters} using bosonization.

However, in order to apply bosonization one has to restrict the elements $g\in \mr N_{G}(\mf h)\subset G$ to the Cartan normalizer, and this will be the key object in our definition of the twist fields.
In particular, bosonization means that one considers the space $\mathcal{H}_{\dot{g}}$ as a sum of twisted representations of the Heisenberg algebra $\widehat{\mathfrak{h}}$. These representations depend not only on the elements $g\in \mr N_{G}(\mf h)$, but also on additional data: eigenvalues of the zero modes of the $g$-invariant part of $\widehat{\mathfrak{h}}$. This extra data, below to be called $r$-charges following \cite{GMtw}, has discrete freedom, since only the exponents of such eigenvalues are specified by $g$. We also denote below the most refined data as $\ddot{g}=(g,\bs r)$.

Finally, we consider the twisted representations of the Heisenberg algebra as (the twist-field) representations of the W-algebra, already \emph{untwisted} since the W-algebra generators are $G$-invariant.

\begin{definition}
We call the vertex operator $\mc V_{\ddot g} = \mc O_{\ddot g}$ a \emph{twist field}
when $g$ lies in the normalizer of Cartan $\mf{h}\subset\mf{g}$, i.e. $g\in\mr N_{G}(\mf h)$ iff
\eq{
g\mf{h}g^{-1}=\mf{h}
\label{conjNC}
}
\end{definition}
Such elements
also preserve the Heisenberg subalgebra
$\widehat{\mathfrak{h}}=\langle J_1(z),\ldots,J_{\rank\ \mf{g}}(z)\rangle\subset\widehat{\mf{g}}_1$
\eq{
	g\widehat{\mathfrak{h}}g^{-1}=\widehat{\mathfrak{h}}
}
If different elements $g_1, g_2\in{\rm N}_G(\mf h)$ are conjugated in $G$: $g_1=\Omega g_2\Omega^{-1}$, then such conjugation identifies $g_1$ and $g_2$ twistings of the fermionic algebras. This conjugation also induces one-to-one correspondence between the set of $\dot{g}_1$ and the set of $\dot{g}_2$,
and maps the twisted representation $\mathcal{H}_{\dot g_1}$ to $\mathcal{H}_{\dot g_2}$.

More formally, if we denote corresponding representations by $T_{\dot g}(\hat{\mathfrak{g}})\colon \hat{\mathfrak{g}}
\rightarrow {\rm End}\mathcal{H}_{\dot g}$, and the action of $\Omega$ by $\varOmega_{12} \colon \mathcal{H}_{\dot g_1} \rightarrow \mathcal{H}_{\dot g_2}$,
then we have
$\varOmega_{12} T_{\dot g_1}(J(z)) \varOmega_{12}^{-1}=T_{\dot g_2}(J(z)^\Omega).$ Note, that twisted representations of KM algebra $\mathcal{H}_{\dot g_1}$,
$\mathcal{H}_{\dot g_2}$ are not isomorphic due to appearance of conjugation by $\Omega$: the current $J(z)$ can have different monodromies
in $\mathcal{H}_{\dot g_1}$ and $\mathcal{H}_{\dot g_2}$.
But the corresponding representations of W-algebra become equivalent (up to external automorphism in case of $SO(2N)$), see details in Sect.~\ref{ssec:W:decom}
\footnote{One can compare this conjugation and additional data in $\dot{g}, \ddot{g}$ with the description of the representations of
$G$-invariant part of vertex algebra in case of the finite group $G$ in \cite{DVVV}.}.

If $g_1=\Omega g_2\Omega^{-1}$, with $\Omega \in \mr N_{G}(\mf h)$, then the conjugation by $\Omega$  preserves $\widehat{\mathfrak{h}}$. Therefore this conjugation induces the transformation of  twisted representations of $\widehat{\mathfrak{h}}$ in $\mathcal{H}_{\dot{g}_1}$ into twisted representations of $\widehat{\mathfrak{h}}$ in $\mathcal{H}_{\dot{g}_2}$, and induces one-to-one correspondence between the sets of $\ddot{g}_1$ and $\ddot{g}_2$. Hence we have an action of $\mr N_{G}(\mf h)$ on the set of $\ddot{g}$.

Below we describe the structure of the Cartan normalizers $\mr N_{GL(N)}(\mf h)$ and
$\mr N_{O(n)}(\mf h)$ and specify notations $\dot{g}$ and $\ddot{g}$ in these cases
 explicitly. We also describe the representatives of all the orbits in the set of $\ddot{g}$ under the action of $\mr N_{G}(\mf h)$.

\subsection{Cartan normalizers}

\paragraph{Structure of the Cartan normalizer for $\mf gl(N)$.}
Let us choose the Cartan subalgebra in a standard way $\mathfrak{h} \supset {\rm diag}(x_1,\ldots,x_N)$, so that the conjugation \rf{conjNC} can only permute the eigenvalues. Therefore we conclude
that
\eq{
g=s\cdot(\lambda_1,\ldots,\lambda_N) \in S_N\ltimes\left(\mathbb C^\times\right)^N = \mr N_{GL(N)}(\mf h)
}
or $g$ is just a quasipermutation.

Now let us find the conjugacy classes in this group. Any element of $\mr N_{GL(N)}(\mf h)$ has the form
$g=(c_1\ldots c_k,(\lambda_1,\ldots,\lambda_N))$, where $c_i$ are the cyclic permutations --- their only parameters are lengths $l_j=l(c_j)$.
It is enough to consider just a single cycle of the length $l=l(c)$
\eq{
g=(c,(\lambda_1,\ldots,\lambda_l))
}
since any $g$ can be decomposed into a product of such elements. Conjugation of this element by diagonal matrix
is given by
\eq{
(1,(\mu_1,\ldots,\mu_l))\cdot(c,(\lambda_1,\ldots,\lambda_l))\cdot(1,(\mu_1,\ldots,\mu_l))^{-1}=
\\
= (c,(\lambda_1\frac{\mu_1}{\mu_2},
\lambda_2\frac{\mu_2}{\mu_3},\ldots,\lambda_l\frac{\mu_l}{\mu_1}))
}
Therefore one can always adjust $\{\mu_i\}$ to replace all $\{\lambda_i\}$ by the same number, e.g.
to put $\lambda_i\mapsto \bar{\lambda}=\prod_{i=1}^l \lambda_i^{1/l}=e^{\pi i\frac{l-1}l}e^{2\pi ir}$. These ``averaged over a cycle''
parameters have been called as $r$-charges in \cite{GMtw}.
Hence, all elements of $g\in\mr N_{GL(N)}(\mf h)$ can be conjugated to the products over the cycles
\eq{
\left[g\right]\sim \prod_{j=1}^K [l_j,e^{i\pi\frac{1-l_j}{l_j}}\bar{\lambda}_j] = \prod_{j=1}^K [l_j,e^{2\pi ir_j}]
\label{glNg}
}
Transformation $[l,e^{2\pi ir}]$ acts like follows:
\eq{
\psi_{\alpha}^*\mapsto e^{i\pi\frac{l-1}l}e^{2i\pi r}\psi_{\alpha+1}^*,\quad\psi_{\alpha}\mapsto e^{i\pi\frac{1-l}l}e^{-2i\pi r}\psi_{\alpha+1}\,,
}
where $\psi_{\alpha+l}=\psi_\alpha$, and we have included the extra factor $e^{i\pi\frac{l-1}l}$ into the definition of transformation in order to have simple formula
\eq{
\det[l,e^{2\pi ir}]=e^{2i\pi rl}
}
and to simplify the identification between $r$ and $U(1)$ charge in Appendix~\ref{app:gln}.

In this case $\dot g$ is just a pair $(g,\tr\log g)$, the value of $\tr\log g$ is defined up to $2\pi i \mathbb{Z}$, and this freedom corresponds to the representation of $\Gamma=U(1)$ mentioned above. Element $\ddot g$ contains information about all $r$-charges  $\ddot g=(g,\bs r)$, for given $g$ the $r$-charge $r_j$ is defined by $g$ up to the shift by $\frac1{l_j}\mathbb{Z}$. If $\ddot g_1$ and $\ddot g_2$ correspond to the same $\dot{g}$, then the corresponding $r$-charges differ by the shift in the certain lattice, see the character formula \rf{eq:chigln} in
Sect.~\ref{sec:characters}.

\paragraph{Structure of $\mr N_{O(n)}(\mf h)$.}

Using complexification of fermions \rf{compexif} we rewrite the quadratic form $ds^2=\Sum_{i=1}^n d\Psi_i^2$
 as $ds^2=\Sum_{\alpha=1}^N d\psi_\alpha^* d\psi_\alpha + d\Psi^2=\Sum_{\alpha=1}^N d\psi_{-\alpha} d\psi_\alpha + d\Psi^2$ (the last term is present only for the
$B_N$-series). In this basis the  $\mf{so}(n)$ algebra (algebra, preserving this form) becomes
just the algebra of matrices, which are antisymmetric under the reflection w.r.t. the anti-diagonal. In particular, the Cartan elements are given by
\eq{
\mathfrak{h}\ni {\rm diag}(x_1,\ldots,x_N,0,-x_N,\ldots,-x_1)
\label{Cartson}
}
for $B_N$-series (and for the $D_N$-series just $0$ in the middle should be removed).
The action of an element from $\mr{N}_{O(n)}(\mf h)$ should preserve the chosen quadratic form,
and, when acting on the diagonal matrix \rf{Cartson}, it can either permute some eigenvalues, also doing it
simultaneously in the both blocks, or interchange $x_\alpha$ with $-x_\alpha$ (the same as to change the sign of $x_\alpha$).
It is defined in this way up to a subgroup of diagonal matrices themselves. In other words
\eqs{
\mr N_{O(2N)}(\mf h)&=S_N\ltimes(\mathbb Z/2\mathbb Z)^N\ltimes(\mathbb C^\times)^N\\
\mr N_{O(2N+1)}(\mf h)&=\mr N_{O(2N)}(\mf h)\times\mathbb Z/2\mathbb Z
}
where the generator of the last factor $\mathbb Z/2\mathbb Z$ changes the sign of the extra fermion $\Psi$.
This triple $(s,\vec{\sigma},\vec{\lambda})\in{\rm N}_{O(n)}(\mf h)$, with $s\in S_N$, $\sigma_\alpha\in \mathbb Z/2\mathbb Z$ and $\lambda_\alpha\in \mathbb C^\times$, is embedded into $O(n)$ as follows
\eqs{
S_N:\ (\{\alpha\mapsto s(\alpha)\},1,1)&=\{\psi_\alpha\mapsto \psi_{s(\alpha)};\ \psi^*_{\alpha}\mapsto \psi^*_{s(\alpha)}\}\\
(\mathbb Z/2\mathbb Z)^N:\ (1,\vec\sigma,1)&=\{\psi_\alpha\mapsto \psi_{\sigma_\alpha\alpha}\}\\
(\mathbb C^\times)^N:\ (1,1,\vec\lambda)&=\{\psi_\alpha\mapsto \lambda_\alpha\psi_\alpha;\
\psi^*_{\alpha}\mapsto \lambda_\alpha^{-1}\psi^*_{\alpha}\}
}
and in these formulas $\psi_{-\alpha}=\psi^*_\alpha$ and $\psi^*_{-\alpha}=\psi_{\alpha}$ is again implied. The structure of the actions in the semidirect product has the obvious form:
\be
\vec \sigma: \lambda_\alpha\mapsto\lambda_\alpha^{\sigma_\alpha},\ \ \ \
s: (\sigma_\alpha,\lambda_\alpha) \mapsto (\sigma_{s(\alpha)},\lambda_{s(\alpha)})
\ee
Notice that the normalizer of Cartan in $SO(n)$
\eq{
\mr N_{SO(n)}(\mf h)=SO(n)\cap\mr N_{O(n)}(\mf h)
}
is distinguished by condition that $\prod_{\alpha=1}^N\sigma_\alpha=1$, and the Weyl group is given as the factor of this normalizer by the Cartan torus
\eq{
\mathrm{W}(\mf{so}(n))=\mr N_{SO(n)}(\mf h)/H
}
Consider now the conjugacy classes in $\mr N_{O(n)}(\mf h)$. First, conjugating an
arbitrary element  $(s,\vec\sigma,\vec\lambda)$ by  permutations, we reduce the problem again to the case when $s=c$ is just a single cycle.
Then one can further conjugate this element by  $(\mathbb Z/2\mathbb Z)^N$:
\eq{
(1,\vec\epsilon,1)\cdot(c,\vec\sigma,\bullet)\cdot(1,\vec\epsilon,1)^{-1}\mapsto (c,(\sigma_1\cdot\epsilon_1\epsilon_2,\sigma_2\epsilon_2\epsilon_3,\ldots,\sigma_N\cdot
\epsilon_1\epsilon_N),\bullet)
}
and solving equations for $\{\epsilon_\alpha\}$, remove all $\sigma_\alpha=-1$, except for, maybe, one.
Hence:

\begin{itemize}
  \item For $\sigma=(1,\ldots,1)$ the situation is the same as in $\mf{gl}(N)$ case: we can transform $\vec\lambda$ to $(\lambda,\ldots,\lambda)$. These conjugacy classes are therefore the same
(but denoted by $[l,\lambda]_+$):
\eq{
(c,1,\vec\lambda)\sim[l(c),e^{i\pi\frac{1-l}l}\prod\lambda_i^{1/l(c)}]_+
}

  \item For, say, $\sigma=(-1,1,\ldots,1)$ let us conjugate this element by $(1,1,\vec \mu)$:
\eq{
(1,1,\vec \mu)(c,(-1,1,\ldots1),\vec\lambda)(1,1,\vec \mu)^{-1}=(c,(-1,1,\ldots,1),\vec\lambda')\\
\vec\lambda'=(\lambda_1\mu_1\mu_2^{-1},\lambda_2\mu_2\mu_3^{-1},\ldots,\lambda_{l-1}\mu_{l-1}\mu_l^{-1},
\lambda_l\mu_l\mu_1)
}
In contrast to the previous case, here one can put all $\lambda_i'=1$, since one can put first $\mu_1^2=\prod_i\lambda_i^{-1}$, and then solve $N-1$ equations
$\mu_{i+1}=\lambda_i\mu_i$ not being restricted by any boundary conditions. It means that
\eq{
(c,(-1,1,\ldots,1),\vec\lambda)\sim[l(c)]_-
}
\end{itemize}

Therefore we can formulate:
\begin{lemma} One gets for the conjugacy classes
\eqs{
\mr N_{O(2N)}(\mf h)&: g\sim\prod_{j=1}^K[l_j,\lambda_j]_+\cdot\prod_{j=1}^{K'}[l_j]_-\\
\mr N_{O(2N+1)}(\mf h)&: g\sim[\epsilon]\cdot\prod_{j=1}^K[l_j,\lambda_j]_+\cdot\prod_{j=1}^{K'}[l_j]_-
\label{olemma}
}
\end{lemma}
and now we are ready to describe the twist fields in detail.

As in the $GL(N)$ case, to be precise one should add explicit values of the $r$-charges, i.e. to consider the pairs $\ddot g=(g,\bs r)$, where $\lambda_j=e^{2\pi i r_j}$.
Moreover, for even $n=2N$ it is also useful to introduce
$\dot g=(g,\widetilde{\bs r})$, where $\widetilde{\bs r}\in \mathbb{R}^K/Q_{D_K}$ are defined up to addition of the vectors from $D_K$ root lattice:
$\bs r\sim \widetilde{\bs r} \mod Q_{D_K}$~\footnote{The vector $\bs r$ is determined by given $g$ up to the lattice $\mathbb Z^K$, (explicitly up to the shift $l_ir_i\mapsto l_ir_i+n_i$ which are due to the logarithms). The lattice $Q_{D_K}\subset \mathbb{Z}^K$ is defined by $\Sum n_i\in2\mathbb Z$. Therefore, there are two elements $\dot g$, corresponding to given $g$, since $|\mathbb Z^K/Q_{D_K}|=2$. These two choices of $r$ exactly correspond to two subspaces with given fermion number $(-1)^F$ in the fermion representations, the group $\Gamma=\mathbb{Z}/2\mathbb{Z}$ acts by this fermion number.}. For odd $n$ we just have $\dot g=g$.

\subsection{Twist fields and bosonization for $\mf{gl}(N)$}

Take an element \rf{glNg},
whose action on fermions (in the fundamental and antifundamental representations), say for a single cycle, is
\eq{
g:\ (\psi_\alpha^*(z), \psi_\alpha(z))\mapsto (e^{i\pi\frac{l-1}l}e^{2\pi ir}\psi_{\alpha+1}^*(z),
e^{i\pi\frac{1-l}l}e^{-2\pi ir}\psi_{\alpha+1}(z)),\ \ \ \ \mod l
\label{bcl}}
while the corresponding (adjoint) action on the Cartan is just
\eq{
g_{\rm Adj}: J_{\alpha}(z)\mapsto J_{\alpha+1}(z),\ \ \ \ \mod l
}
Such formulas have a simple geometric interpretation \cite{Knizhnik}: there is the branched cover
in the vicinity of the point $z=0$ given by $\xi^l=z$, so that all these (fermionic and bosonic)
fields are actually defined on different sheets $\xi^{(\alpha)} = z^{1/l}e^{2\pi i\alpha/l}$ of the cover:
\eq{
\psi_\alpha^*(z)\sqrt{dz} =\tilde\psi^*(\xi^{(\alpha)})\sqrt{d\xi^{(\alpha)}},\ \ \ \
\psi_\alpha(z)\sqrt{dz}
=\tilde\psi(\xi^{(\alpha)})\sqrt{d\xi^{(\alpha)}}
\\
J_\alpha(z)dz=J(\xi^{(\alpha)})dz=\tilde J(\xi^{(\alpha)})d\xi^{(\alpha)}
}
Using these formulas one can write down expansions for the fields on the cover, whose OPE's locally are given by
\eq{
\tilde\psi^*(\xi)\tilde\psi(\xi')=\frac1{\xi-\xi'} + reg.,\ \ \ \ \
\tilde J(\xi)\tilde J(\xi')=\frac1{(\xi-\xi')^2}+ reg.
}
Now one writes the following formulas for the mode expansion of fermions, which are already twisted on the covering curve by $e^{2\pi i\sigma}$:
\eq{
\psi(z)=\sqrt{\frac{d\xi}{dz} }\tilde\psi(\xi)=\frac{z^{\frac1{2l}-\frac12}}{\sqrt{l}}
\Sum_{p\in\mathbb Z+\frac12}\frac{\psi_p}{\xi^{p+\frac12+\sigma}}
=\frac1{\sqrt{l}}\Sum_{p\in\mathbb Z+\frac12}\frac{\psi_p}{z^{\frac12+\frac1l (p+\sigma)}}\\
\psi^*(z)=\sqrt{\frac{d\xi}{dz} }\tilde\psi^*(\xi)=\frac{z^{\frac1{2l}-\frac12}}{\sqrt{l}}
\Sum_{p\in\mathbb Z+\frac12}\frac{\psi_p^*}{\xi^{p+\frac12-\sigma}}
=\frac1{\sqrt{l}}\Sum_{p\in\mathbb Z+\frac12}\frac{\psi_p^*}{z^{\frac12+\frac1l (p-\sigma)}}
}
Due to \rf{bcl} one should have $\psi^*(e^{2\pi i l}z)=(-1)^{l-1}e^{2\pi ilr}\psi^*(z)$ and $\psi(e^{2\pi i l}z)=(-1)^{l-1}e^{-2\pi ilr}\psi(z)$,
therefore one can take $\sigma=lr$, so that:
\eq{
\psi(z)=\frac1{\sqrt{l}}\Sum_{p\in\mathbb Z+\frac12}\frac{\psi_{p}}{z^{\frac1l(\frac12+p)+r+\frac{l-1}{2l}}},\quad
\quad\psi^*(z)=\frac1{\sqrt{l}}\Sum_{p\in\mathbb Z+\frac12}\frac{\psi_{p}^*}{z^{\frac1l(\frac12+p+l-1)-r-\frac{l-1}{2l}}},
\\
\left\{\psi_{p},\psi_{p'}^*\right\} = \delta_{p+p',0}
\label{psitwist}
}
or the mode expansion is shifted by the $r$-charges, corresponding to given cycles.

The same procedure gives for the twisted bosons
\eq{
J(z)=\frac1l z^{\frac1l-1}\tilde J(\xi)=\frac1l z^{\frac1l-1}\Sum_{n\in\mathbb Z}\frac{J_{n/l}}{\xi^{n+1}}
=\frac1l z^{\frac1l-1}\Sum_{n\in\mathbb Z}\frac{J_{n/l}}{z^{\frac1l(n+1)}}=\frac1l\Sum_{n\in\mathbb Z}\frac{J_{n/l}}{z^{\frac1l n+1}}
\label{Jtwist}
}
with the commutation relations between their modes being
\eq{
\left[J_{n/l},J_{m/l}\right]=n\delta_{n+m,0}\ \ \ \ \ n,m\in \mathbb{Z}
}
These twisted bosons provide one of the convenient languages for the twist field representations. The other
one is provided by bosonization of the constituent fermions with the fixed fractional parts of the power expansions in \rf{psitwist}
\be
\psi(z) = \frac1{\sqrt{l}}\Sum_{a \in\mathbb Z/l\mathbb{Z}}\psi_{(a)}(z),\ \ \ \
\psi_{(a)}(e^{2\pi i}z) = e^{-i\pi\frac{1-l}l-2\pi ir - 2\pi i\frac{a}{l}}\psi_{(a)}(z),
\\
\psi^*(z) = \frac1{\sqrt{l}}\Sum_{a \in\mathbb Z/l\mathbb{Z}}\psi^*_{(a)}(z),\ \ \ \
\psi^*_{(a)}(e^{2\pi i}z) = e^{i\pi\frac{1-l}l+2\pi ir + 2\pi i\frac{a}{l}}\psi^*_{(a)}(z)
\ee
The corresponding bosons (see \rf{bosonization01} in Appendix)
\be
I_{(a)}(z) = \left(\psi^*_{(a)}(z)\psi_{(a)}(z)\right) = \sum_{n\in \mathbb{Z}} \frac{I^{(a)}_n}{z^{n+1}}
+ \frac1z \left(r+ \frac{a}{l}+\frac{1-l}{2l}\right)
\ee
always have an integer mode expansion.

\subsection{Twist fields and bosonization for $\mf{so}(n)$} \label{ssec:bos:so(n)}

Let us mention first that there is a difference between the groups $\mr N_{O(n)}(\mf h)$ and $\mr N_{SO(n)}(\mf h)$, since the action of the first
one can also map
$V(z)\mapsto -V(z)$, so that one of the generators of the W-algebra $V(e^{2\pi i}z)=-V(z)$ becomes a Ramond field, and we allow this extra minus sign below~\footnote{Note that for $n=2N$ the action of $\mr N_{SO(2N)}(\mf h)$ on $\mathfrak{h}$ is given by the Weyl group action, but an extra element from $\mr N_{O(2N)}(\mf h)$ gives external (diagram) automorphism. Corresponding twisted representations could be viewed as a representation of twisted affine Lie algebra $D_N^{(2)}$.}.

In addition to the conjugacy classes $[l,\lambda]_+$, similar to those of $\mf{gl}(N)$, we now also have to study $[l]_-$'s. First one has to identify the action of $\mr N_{O(n)}(\mf h)$ on the fermions, where just by definition:
\eq{
\sigma_\alpha=-1:\ \ \  (1,\vec\sigma,1):\psi_\alpha\mapsto\psi^*_{\alpha}
}
This means that the element of our interest is the complete cycle
\eq{
\left[l\right]_-:\psi_1\mapsto\psi_2\mapsto\ldots\mapsto\psi_l\mapsto\psi_1^*\mapsto\ldots
\mapsto\psi_l^*\mapsto\psi_1
\label{psimon}
}
Therefore $2N$ complex fermions can be realized as a pushforward of a single real fermion $\eta(\xi)$, living on a $2l$-sheeted branched cover
\eq{
\psi_\alpha(z)\sqrt{dz}
=\eta(\xi^{(\alpha)})\sqrt{d\xi}\\
\psi^*_\alpha(z)\sqrt{dz}
=\eta(\xi^{(l+\alpha)})\sqrt{d\xi}
}
Here the branched cover $z=\xi^{2l}$ can be realized as a sequence of two covers $\pi_2: \xi\mapsto \zeta=\xi^2$ and $\pi_l: \zeta\mapsto \zeta^l=z$, and it leads to trickier global construction of
the exact conformal blocks, see Sect.~\ref{section:confblock} below.

An important fact is that there is an element $\mc\sigma\in (\mr{N}_{O(n)}(\mf h)/H)$ in the center of this group
\eq{
\mc\sigma=(1,(-1,-1,\ldots,-1))
}
which generates the global automorphism of the cover of order two, which is continued to the global automorphism of the algebraic curve
during the consideration of exact conformal blocks in Sect.~\ref{section:confblock}. It acts locally by $\xi\mapsto-\xi$. Using this element one can
write the OPE of $\eta(\xi)$ in the form:
\eq{
\eta(\xi)\eta(\sigma(\xi'))=\frac1{\xi-\xi'}+reg.
}
Now the analytic structure of this field can be obtained
\eq{
\psi(z)= \sqrt{\frac{d\xi}{dz}}\eta(\xi) = \frac{z^{\frac1{4l}-\frac12}}{\sqrt{2l}}\Sum_{p\in\mathbb Z+\frac12}\frac{\eta_{p+\frac12}}{z^{\frac1{2l}(p+\frac12+\sigma)}} =
\frac{1}{\sqrt{2l}}\Sum_{p\in\mathbb Z+\frac12}
\frac{\eta_{p+\frac12}}{z^{\frac1{2l}(p+\sigma)+\frac12}}
\\
\psi^*(z)=\psi(e^{2\pi il}z)
\label{etapsi}
}
In order to ensure right monodromies \rf{psimon} for $\psi$, $\psi^*$ one should get powers $\frac1{2l}\mathbb Z$ in the r.h.s., which means
that $\sigma\sim l-\frac12\sim\frac12$,
and $\eta(\xi)$ turns to be a Ramond fermion with the extra ramification
\eq{
\eta(\xi)=\Sum_{n\in\mathbb Z}\frac{\eta_n}{\xi^{n+\frac12}},\ \ \ \ \
\psi(z) = \frac{1}{\sqrt{2l}}\Sum_{n\in\mathbb Z}\frac{\eta_{n}}{z^{\frac{n}{2l}+\frac12}},\ \ \ \
\psi^*(z) = \frac{(-)^l}{\sqrt{2l}}\Sum_{n\in\mathbb Z}\frac{(-)^n\eta_{n}}{z^{\frac{n}{2l}+\frac12}}
\label{psieta}
}

Let us now construct (a twisted!) boson from this fermion by
\eq{
J(z)=\left(\psi^*(z)\psi(z)\right)=\left(\psi(e^{2i\pi l}z)\psi(z)\right)
}
This boson behaves as follows under the continuation around a twist field:
\eq{
J_1\mapsto J_2\mapsto\ldots\mapsto J_l\mapsto -J_1\mapsto\ldots\mapsto -J_l
}
To realize this situation we may take the Ramond boson on the cover in variable $\zeta$:
\eq{
J(z)=\frac{d\zeta}{dz}\sum_{r\in \mathbb{Z}+\frac12}\frac{{J}_{r/l}}{\zeta^{r+1}} =
\frac{z^{\frac1{l}-1}}{l}\sum_{r\in \mathbb{Z}+\frac12}\frac{{J}_{r/l}}
{z^{\frac1{l}(r+1)}}=\frac1{l}\sum_{r\in \mathbb{Z}+\frac12}\frac{{J}_{r/l}}
{z^{\frac{r}{l}+1}}
}
where commutation relations of modes are given by
\eq{
\left[J_{r/l},J_{r'/l}\right]=r\delta_{r+r',0}\ \ \ \ r,r'\in \mathbb{Z} + \half
}
Inverse  bosonization formula for this real fermion looks like
\eq{
\sqrt{z}\psi(z)=\Sum_{n\in\mb Z}\frac{\eta_n}{z^{\frac n{2l}}}=\frac{\sigma_1}{\sqrt 2}e^{i\phi_-(z)}e^{i\phi_+(z)}
}
with the Pauli matrix $\sigma_1 = \left(\begin{array}{cc}
                                          0 & 1 \\
                                          1 & 0
                                        \end{array}\right)
$, and it is discussed in detail in Appendix~\rf{NSR}.

\section{Characters for the twisted modules}
\label{sec:characters}

Now we turn directly to the computation of characters, using bosonization rules. In order to do this one has to apply the following heuristic ``master formula'' for the trace
\eq{
\chi_{\dot g}(q) = \tr_{\mathcal{H}_{\dot g}} q^{L_0}\;``="\;\frac{\chi_{ZM}(q)}{\Prod_k\Prod_{n=1}^\infty(1-q^{\theta_{{\rm Adj},k}(g)+n})}
\label{chi1}
}
over the space $\mc H_{\dot g}$, which is the minimal space closed under the action of both W-algebra and twisted Kac-Moody algebra. For simply-laced cases,
$\mf{gl}(N)$ and
$\mf{so}(2N)$, $\mc H_{\dot g}$ is  the module of corresponding Kac-Moody algebra, whereas in the $\mf{so}(2N+1)$ case it should be entire fermionic Fock module
due to presence of the fermionic W-current. Explicit descriptions of $\mc H_{\dot g}$ are the following: for $\mf{gl}(N)$ it is the subspace
with the fixed total fermionic charge, for $\mf{so}(2N)$ it is the subspace with the fixed parity of total fermionic charge, and for $\mf{so}(2N+1)$ it
is entire space.

Notice that this representation depends on $\dot{g}$ --- not just $g$, and also it contains all representations of W-algebra with
different $\ddot g$ corresponding to the same $\dot g$.

Denominator of \rf{chi1} collects the contributions from the Fock descendants of twisted bosons (parameters $\theta_{{\rm Adj},k}(g)$ are the
eigenvalues of adjoint action of $g$ on the Cartan subalgebra), and the numerator --- contribution of the zero
modes. This formula is heuristic, moreover, in some important cases we also get contribution
from the extra fermion, sometimes it is more informative to
consider super-characters etc. Below we prove the following
\begin{theorem} The characters of twisted representations are given by the formulas \rf{eq:chigln} \rf{chiD1}, \rf{chiD2}, \rf{chiB1}, \rf{chiB2}.
\end{theorem}

\subsection{$\mf{gl}(N)$ twist fields}

To be definite, let us fix an element $g=\Prod_{j=1}^K[l_j,e^{2\pi i r_j}]$ from \rf{glNg} which, according to \rf{bcl}, performs the permutation of
fermions with the simultaneous multiplication by $e^{\pm 2\pi i (r_j+\frac{l_j-1}{2l_j})}$. In this setting $N$ fermions can be bosonized in terms of $K$ twisted bosons (see details in Appendix~\ref{app:gln}), and  here we just present the final formulas
\be
\psi^*_\alpha(z)=\frac{z^{\frac{1-l}{2l}}}{\sqrt{l}}e^{i\phi^{(j)}_-(e^{2\pi i\alpha} z)}e^{i\phi^{(j)}_+(e^{2\pi i \alpha} z)}
e^{Q^{(j)}}(e^{2\pi i\alpha} z)^{\frac1lJ^{(j)}_0}
(-1)^{\Sum_{k<j}J^{(k)}_0}
\\
\psi_\alpha(z)=\frac{z^{\frac{1-l}{2l}}}{\sqrt{l}}e^{-i\phi^{(j)}_-(e^{2\pi i\alpha} z)}e^{-i\phi^{(j)}_+(e^{2\pi i\alpha} z)}
e^{-Q^{(j)}}(e^{2\pi i\alpha} z)^{-\frac1lJ^{(j)}_0}
(-1)^{\Sum_{k<j}J^{(k)}_0}
\label{bosonization02}
\ee
for $\alpha\in \mathbb{Z}/l_j\mathbb{Z}$, labeling the fields within $[l_j]$-cycle.
For the conformal dimension one gets therefore (see \rf{L0twisted}, and computation by alternative methods
in \rf{eq:Deltas1}, \rf{dimD})
\eq{
L_0=\Sum_{j=1}^K\frac{l_j^2-1}{24l_j}+\Sum_{j=1}^K\frac1{2l_i}(J_0^{(j)})^2+\ldots
}
and since we are computing character of the space, obtained by the action of $\widehat{\mf{gl}(N)}_1$, we  have to take into account all
vacua arising after the action of the shift operators $e^{Q^{(i)}-Q^{(j)}}$, i.e. labeled by the $A_{K-1}$ root lattice. Hence, the character \rf{chi1} for this case is given by
\eq{
\chi_{\dot g}(q)=q^{\Sum_{j=1}^K\frac{l_j^2-1}{24l_j}}\frac{\Sum_{n_1+\ldots+n_K=0}
q^{\Sum_{i=1}^K\frac1{2l_i}(r_il_i+n_i)^2}}{
\Prod_{j=1}^K\Prod_{k=1}^\infty(1-q^{k/l_j})}
\label{eq:chigln}
}
In this formula the numerator collects contributions from the highest vectors $\chi_{ZM}$ (they differ by the value of zero modes $J_0^{(i)}$) of the Heisenberg algebras
with generators $J^{(i)}_{n/l_i}$, whereas the denominator contains the contributions from the descendants.

\subsection{$\mf{so}(2N)$ twist fields, $K'=0$}

Consider now the twist fields \rf{olemma} for $g\in\mr N_{O(2N)}(\mf h)$, and take first
$K'=0$, so our twist has no minus-cycles
\eq{
g=\Prod_{j=1}^K[l_j,e^{2\pi i r_j}]_+
}
 The only difference from the previous situation with the $\mf{gl}(N)$ case is that now one also have
extra currents $J_{\alpha\bar\beta}=\psi^*_\alpha(z)\psi^*_\beta(z)$ and $J_{\bar\alpha\beta}=\psi_\alpha(z)\psi_\beta(z)$. It means
that due to bosonization \rf{bosonization01}, \rf{bosonization02} possible charge's shifts now include $e^{\pm Q^{(i)}\pm Q^{(j)}}$,
so the full lattice of the zero-mode charges (one zero mode for each
cycle $[l_i,e^{2\pi i r_i}]_+$) contains all points with
\eq{
\Sum_{i=1}^K n_i\in2\mb Z,\ \ \ \ \{ n_i\}\in \mathbb{Z}^K
}
or is just the root lattice $Q_{D_K}$. After corresponding modification of numerator and the same
contribution of the twisted Heisenberg algebra to denominator, the formula for the character in this case acquires the form
\eq{
\chi_{\dot g}(q)=q^{\Sum_{j=1}^K\frac{l_j^2-1}{24l_j}}\frac{\Sum_{\vec n\in Q_{D_K}}q^{\Sum_{j=1}^K\frac1{2l_j}(n_j+l_jr_j)^2}}{\Prod_{j=1}^K
\Prod_{n=1}^\infty(1-q^{n/l_j})}
\label{chiD1}
}

\subsection{$\mf{so}(2N)$ twist fields, $K'>0$}
Take
\eq{
g=\Prod_{j=1}^K[l_j,e^{2\pi i r_j}]_+\Prod_{j=1}^{K'}[l_j']_-
\label{geno2N}
}
Now we have extra cycles of type $[l'_i]_-$, so we have extra $\eta$-fermions that have to be bosonized in a different way \rf{bosonizationeta}:
\eq{
\eta_i(z)=\frac{z^{-\frac12}}{2\sqrt{l}}e^{i\phi_-(z^{\frac1{2l_i}})}e^{i\phi_+(z^{\frac1{2l_i}})}
(-1)^{\Sum_k J^{(k)}_0}\gamma_i
}
where $\{\gamma_i,\gamma_j\}=2\delta_{ij}$ are gamma-matrices (or generators of the Clifford algebra $\mc{C}l_{K'}(\mathbb C)$) in the smallest possible representation, which
make different fermions anticommuting.
Due to presence of $K'$ cycles of type $[l_i']_-$, the zero-mode $\chi_{ZM}(q)$ generating operators
include now $\gamma_je^{Q^{(i)}}$, which perform integer shifts of $i$-th bosonic zero mode together with the inessential action on
fermionic vacua. So now we do not have to imply that the number of shifts by $e^{Q^{(i)}}$ should be even. Hence, instead of $D_K$-lattice from \rf{chiD1}
the numerator includes now summation over the root lattice $Q_{B_K}$, i.e.
\eq{
\chi_{\dot g}(q)=q^{\Delta^0_g}
\frac{2^{[\frac{K'+1}2]-1}\Sum_{\vec n\in Q_{B_K}}q^{\Sum_{i=1}^K\frac1{2l_i}(n_i+l_ir_i)^2}}{\Prod_{i=1}^K\Prod_{k=1}^\infty(1-q^{k/l_i})
\Prod_{i=1}^{K'}\Prod_{k=0}^\infty(1-q^{(k+\frac12)/l'_i})}
\label{chiD2}
}
where factor $2^{[\frac{K'+1}2]-1}$ corresponds to the dimension of the spinor representation of $\mf so(K')$, generated by $\gamma_i\gamma_j$. Another simple factor $q^{\Delta^0_g}$ contains the minimal
conformal dimension (without contribution of the ``$r$-charges'')
\eq{
\Delta^0_g=\Sum_{i=1}^K\frac{l_i^2-1}{24 l_i}+\Sum_{i=1}^{K'}\frac{2l_i'^2+1}{48 l_i'}
\label{deltag}
}
which will be computed below in \rf{dimD0}, \rf{dimD}. The numerator of \rf{chiD2} contains $K$ contributions from twisted bosons corresponding to plus-cycles,
and $K'$ contributions from twisted Ramond bosons corresponding to minus-cycles.

\subsection{$\mf{so}(2N+1)$ twist fields}

The W-algebra $W(\mf{so}(2N+1))$ contains fermionic operator $V(z)=\Psi_1(z)\ldots\Psi_{2N+1}(z)$,
which cannot be expressed in terms of generators of $\widehat{\mf{so}}(2N+1)_1$ since the latter are all even in fermions. It means that to
construct a module of the W-algebra one should use entire fermionic algebra. Taking into account the fermionic nature of this W-algebra
one can consider $\mathbb{Z}/2\mathbb Z$ graded modules and define two different characters
\eq{
\chi^+(q)=\tr q^{L_0},\ \ \ \ \
\chi^-(q)=\tr (-1)^F q^{L_0}
}
where $F$ is the fermionic number:
\eq{
(-1)^FU_k(z)=U_k(z)(-1)^F,\ \ \ \
(-1)^FV(z)=-V(z)(-1)^F
}
One of the characters vanishes $\chi^-(q)=0$ if at least one fermionic zero mode exists, since each state gets partner with the opposite fermionic parity.
Such fermionic zero modes are always present for the Ramond fermions and $\eta$-fermions, so the only case with non-trivial $\chi^-(q)$ corresponds to
\eq{
g=[1]\Prod_{i=1}^{K}[l_i,e^{2\pi i r_i}]_+
}
In this case our computation works as follows: take bosonization for the $[l]_+$-cycles in terms of $K$ twisted bosons \rf{bosonization01}, \rf{bosonization02}, then the
fermionic operators produce the zero-mode shifts $e^{\pm Q^{(i)}}$ with the fermionic number $F=F^b+F^f=F^b=1$, and the Heisenberg generators $J^{(i)}_{n/l_i}$ with the fermionic number
$F=F^b=0$. Moreover, we also have an extra ``true'' fermion $\Psi(z)$ with $F=F^f=1$.
Therefore the total trace can be computed, separating bosons and fermions, as
\eq{
\chi^-(q)=\tr q^{L_0}(-1)^F=\tr q^{L^{b}_0}(-1)^{F^{b}}\cdot \tr q^{L^{f}_0}(-1)^{F^{f}}
}
where the traces over bosonic and fermionic spaces are given by
\eq{
\tr q^{L^{b}_0}(-1)^{F^{b}}=\frac{\Sum_{n_1,\ldots,n_K\in\mathbb Z}q^{\Sum_{i=1}^K\frac1{2l_i}(n_i+l_ir_i)^2}(-1)^{\Sum_{i=1}^K n_i}}{
\Prod_{i=1}^K\Prod_{n=1}^\infty(1-q^{n/l_i})}
\\
\tr q^{L^{f}_0}(-1)^{F^{f}}=\prod_{n=0}^\infty(1-q^{n+\frac12})
}
Hence, the final answer for this character is given by
\eq{
\chi^-_{\dot g}(q)=q^{\Delta^0_g}\frac{\left(\Sum_{\vec n\in Q_{D_K}}q^{\Sum_{i=1}^K\frac1{2l_i}(n_i+l_ir_i)^2}-
 \Sum_{\vec n\in Q_{D_K'}}q^{\Sum_{i=1}^K\frac1{2l_i}(n_i+l_ir_i)^2}\right)\Prod_{k=0}^\infty(1-q^{k+\frac12})}{
\Prod_{i=1}^K\Prod_{k=1}^\infty(1-q^{k/l_i})}
\label{chiB1}
}
where $D$- and $D'$-lattices are defined in \rf{ADD}.

Let us now turn to the computation of $\chi^+(q)$.
Choose an element from ${\rm N}_{O(2N+1)}(\mf h)$
\eq{
g=[(-1)^{a+1}]\Prod_{i=1}^{K}[l_i,e^{2\pi i r_i}]_+\Prod_{i=1}^{K'}[l'_i]_-
}
where $a=0,1$. The bosonized fermions $e^{i\varphi^{(i)}(z)}$ contain elements $e^{Q^{(i)}}$
generating the $B_K$ root lattice, which together with contribution of the fermionic and Heisenberg modes finally give
\eq{
\chi^+_{\dot g}(q)=q^{\Delta^0_g}\frac{2^{[\frac{K'}2]}\Sum_{\vec n\in Q_{B_K}}q^{\Sum_{i=1}^K\frac1{2l_i}(n_i+l_ir_i)^2}
\Prod_{k=0}^\infty(1+ q^{k+\frac a2})}{
\Prod_{i=1}^K\Prod_{k=1}^\infty(1-q^{k/l_i})\Prod_{i=1}^{K'}\Prod_{k=0}^\infty(1-q^{(k+\frac 12)/l'_i})}
\label{chiB2}}
where
\eq{
\Delta^0_g=\frac{\delta_{a,0}}{16}+\Sum_{i=1}^K\frac{l_i^2-1}{24 l_i}+\Sum_{i=1}^{K'}\frac{2l_i'^2+1}{48 l_i'}
}
Here the only new part, comparing to the $D_N$-case, is an extra factor
\eq{
\chi_f(q)=q^{\frac{\delta_{a,0}}{16}}\prod_{k=0}^\infty(1+q^{\frac a2+k})
}
corresponding to ($R$ or $NS$) fermionic contribution.

\subsection{Character identities
\label{ss:Chariden}}

In Sect.~\ref{ss:twists} we have classified the twist fields by conjugacy classes in ${\rm N}_G(\mf h)$ (more precisely, by the orbits of ${\rm N}_G(\mf h)$ on the set of $\ddot{g}$).
However, it is possible that two different elements $g_1, g_2\in{\rm N}_G(\mf h)$ in the normalizer of Cartan are nevertheless conjugated in $G$: $g_1=\Omega g_2\Omega^{-1}$. As was explained in sec. \ref{ssec:twistmain}, conjugation by $\Omega$ maps twisted representation $\mathcal{H}_{\dot g_1}$ to $\mathcal{H}_{\dot g_2}$. The explicit formula is
$\varOmega_{12} T_{\dot g_1}(J(z)) \varOmega_{12}^{-1}=T_{\dot g_2}(J(z)^\Omega).$
Since  $\psi^*_\alpha(z)^\Omega=\Omega_{\alpha\beta}\psi_\beta^*(z)$, $\psi_\alpha(z)^\Omega=\Omega^{-1}_{\beta\alpha}\psi_\beta(z)$,  grading operator
is invariant $L_0^\Omega=L_0$, and we have

\begin{theorem}
If $\dot g_1\sim \dot g_2$ in $G$ for different $g_1, g_2\in{\rm N}_{G}(\mf h)$, then $\chi_{\dot g_1}(q)=\chi_{\dot g_2}(q)$.
\end{theorem}
This theorem is sometimes an origin of non-trivial identities and product formulas for the lattice theta-functions, and below we examine such examples.

\paragraph{$\mf{gl}(N)$ case.} Here any element is conjugated to a product of cycles of length one
(see the exact definition of shifted $r$-charge in \rf{glNg}):
\eq{
\left[l,e^{2\pi i r}\right]\sim\prod\limits_{j=0}^{l-1}[1,e^{2\pi i v_j^{(l,r)}}]\,,
\label{glsim}
}
where $v_j^{(l,r)}=r+\frac{1-l+2j}{2l}$. One gets therefore an identity
\eq{
\frac{\Sum_{k_1+\ldots+k_N=0}q^{\frac12\Sum_{i=1}^N(v_i+k_i)^2}}{\eta(q)^N}=
\frac{\Sum_{n_1+\ldots+n_K=0}q^{\Sum_{i=1}^K\frac1{2l_i}(n_i+l_ir_i)^2}}{\Prod_{i=1}^K\eta(q^{1/l_i})}
\label{glNidt}
}
where~\footnote{Notation $\vec v\oplus \vec u$ means $(v_1,\ldots,v_k)\oplus(u_1,\ldots u_m)=(v_1,\ldots,v_k,u_1,\ldots,u_m)$.}
$v=v^{(l_1,r_1)}\oplus\ldots\oplus v^{(l_K,r_K)}$. All conformal dimensions for vanishing $r$-charges are conveniently absorbed by the Dedekind
eta-functions $\eta(q)=q^{1/24}\prod_{n=1}^\infty(1-q^n)$.

This equality of characters can be checked by direct computation, see \rf{identity1} in Appendix~\ref{app:thetaid} for $S=\{0\}$. For a single cycle $K=1$ this gives a product
formula for the lattice $A_{N-1}$-theta function \rf{ANprod}, which for $N=2$
\be
\frac{q^{\frac1{16}}}{\prod_{k\geq 0}\left(1-q^{k+1/2}\right)} =
\frac{\sum_{n\in \mathbb{Z}} q^{(n+1/4)^2}}{\prod_{n>0}(1-q^n)}
\ee
was known yet to Gauss and has been originally used by Al.~Zamolodchikov in the context of twist-field representations of the Virasoro algebra.

\paragraph{$\mf{so}(2N)$ case.}
For the conjugacy classes of the first type we have again \rf{glsim}, or
\eq{
\left[l,e^{2\pi i r}\right]_+\sim\prod\limits_{j=0}^{l-1}[1,e^{2\pi i v_j^{(l,r)}}]_+
}
which leads to very similar identities to the $\mathfrak{gl}(N)$-case. For example, one can easily
rederive the product formula \cite{Macdonald:1972} for the $D$-lattice theta function
\be
\label{chD}
\sum_{\vec{n}\in Q_{D_N}}q^{\frac12 \left(\vec{n}+\vec{v}'\right)^2} = \Theta_{D_N}\left(\vec{v}'|q\right)
= \frac{\eta(q)^{N+1}\eta(q^{1/(N-1)})}{\eta(q^{1/2})\eta(q^{1/2(N-1)})}
\ee
for $\vec{v}'=\frac{\vec{\rho}}{h}$, where the structure of product in the r.h.s. again comes from the characteristic polynomial of the Coxeter element of the Weyl group $\texttt{W}(D_N)$. Here $h=2(N-1)$ is the Coxeter number, and $\vec \rho=(N-1,N-2,\ldots,1,0)$ is the Weyl vector, corresponding
to the twist field with dimension $\Delta = \Delta^0 = \frac{N(2N-1)}{48(N-1)}$, and the easiest way
to derive \rf{chD} is to use \rf{identity2} from Appendix~\ref{app:thetaid}.

For another type of the conjugacy classes $[l]_-$, the situation is trickier. The corresponding $\eta$-fermion
\eq{
\eta(z)=z^{-\frac12}\Sum_{k\in\mathbb Z}\frac{\eta_k}{z^{\frac k{2l}}}
}
can be separated into the parts with fixed monodromies around zero:
\eq{
\eta_{(a)}=z^{-\frac12}\Sum_{k\in\mathbb Z}\frac{\eta_{a+2l\cdot k}}{z^{\frac a{2l}+k}}\,,
}
so that the only non-trivial OPE is between $\eta_{(a)}$ and $\eta_{(2l-a)}$. In particular, $\eta_{(0)}$ and $\eta_{(l)}$ are
self-conjugated Ramond (R) and Neveu-Schwarz (NS) fermions, which can be combined into the new $\bar\eta$ fermion, whereas all other components can be considered as charged twisted fermions $\bar\psi, \bar\psi^*$:
\be
\bar\psi_{(a)}(z)=\eta_{(a)}(z),\ \ \ \
\bar\psi^*(z)=\eta_{(2l-a)}(z),\ \ \ \ a=1,\ldots,l-1
\\
\bar\eta(z)=\eta_{(0)}(z)+\eta_{(l)}(z)
\ee
Therefore one gets equivalence
\be
\left[l\right]_-\sim[1]_-\cdot\Prod_{j=1}^{l-1}[1,e^{2\pi i \tilde v^{(l)}_j}]\,,
\ee
where $\tilde v^{(l)}_j=\frac j{2l}$.

Moreover, if we take the  product of two cycles $[1]_-$, then we can combine a pair of $R$-fermions and a pair
of $NS$-fermions into two complex fermions with charges $0$ and $\frac12$, therefore
\eq{
\left[1\right]_-[1]_-\sim[1,1]_+[1,-1]_+
}
This means literally that a pair of $\eta$-fermions is equivalent to two charged bosons: one with charge $v=0$ and another one with charge $v=\frac12$.
Equivalence between these two representations leads to the simple identity \rf{JacobiProd1}, \rf{JacobiProd2}:
\eq{
\frac{2q^{\frac18}}{\Prod_{n=1}^\infty(1-q^{n+\frac12})^2}=\frac{\Sum_{k,n\in\mathbb Z}q^{\frac12n^2+\frac12(k+\frac12)^2}}{\Prod_{n=1}^\infty(1-q^n)^2}
}
Using this identity we can remove a pair of $[1]_-$ cycles from \rf{chiD2} shifting $K'\mapsto K'-2$, and add two more directions to the lattice of charges
$B_K\mapsto B_{K+2}$ with corresponding $r$-charges $0$ and $\frac12$.

\paragraph{$\mf{so}(2N)$ case, $K'=0$.}
We have the consequence of identity \rf{identity1} for the case  $S=2\mathbb Z$:
\eq{
\Sum_{\vec k\in Q_{D_N}}q^{\frac12\Sum_{i=1}^N(v_i+k_i)^2}=
\Prod_{i=1}^K\frac{\eta(q)^{l_i}}{\eta(q^{\frac1{l_i}})}\cdot\Sum_{\vec n\in Q_{D_K}}q^{\Sum_{i=1}^K\frac1{2l_i}(n_i+l_ir_i)^2}
}
where $v=v^{(l_1,r_1)}\oplus\ldots\oplus v^{(l_K,r_K)}$.
\paragraph{$\mf{so}(2N)$ case, $K'>0$; $\mf{so}(2N+1)$, $K'>0$.}
In these cases everything can be expressed in factorized form using \rf{identity2} and checked explicitly, so these cases are not very interesting.

\paragraph{$\mf{so}(2N+1)$ case, NS fermion.}
Here in addition to all identities that we had in the $\mf{so}(2N)$ case, one has two more identities  appearing due to the fact that we
can combine the $NS$ (or $R$) fermion with a pair of $NS, R$ fermions to get one complex fermion with twist $0$ (or twist $\frac12$) and one R-fermion
(or NS-fermion).  Thus
\eqs{
\left[1\right]\cdot[1]_-&\sim [-1]\cdot[1,1]_+\\
[-1]\cdot[1]_-&\sim [1]\cdot[1,-1]_+\\
}
Due to these relations in the cases with $K'\neq 0$ one can always transform any character with the $NS$ fermion to a character with the $R$ fermion, and vice versa.

\subsection{Twist field representations and modules of W-algebras}
\label{ssec:W:decom}

By definition, all our twisted representations of the Kac-Moody algebra are twist-field representations of the W-algebra. As was explained in previous section, if $g_1=\Omega g_2\Omega^{-1}$, then conjugation by $\Omega$ transforms the $g_1$-twisted representation to the $g_2$-twisted representation. Moreover, such conjugation transforms the W-algebra generators expressed through the $g_1$-twisted fermions to those expressed through the $g_2$-twisted fermions with a single exception: if $\Omega\in O(n)\setminus SO(n)$, the conjugation by $\Omega$ changes the sign of the last generator $V(z)$, see \eqref{eq:WgenBD}. For odd $n$ this is equivalent to the action of the operator $(-1)^F$ on representations of W-algebra, but for even $n$ this is an external automorphism of the W-algebra (coming from external automorphism of $D_N$).

Therefore, the representations of W-algebra corresponding to conjugated $g$-twists are isomorphic, except for the case when $\Omega\in O(2N)\setminus SO(2N)$ --- where only external automorphism of W-algebra maps one representation to another. The last detail is not crucial if twist $g$ commutes with a certain element of $O(2N)\setminus SO(2N)$ --- in this case any conjugation $\Omega$ can be reduced to the conjugation
by $SO(2N)$. This happens when $g$ belongs to the class $\prod_{j=1}^n[1,e^{2\pi i v_j}]_+$ with some $v_k=0$ or $v_k=\frac12$: for example, it can be
obtained from a pair of minus-cycles, or from some plus-cycle with the fine-tuned $r$-charge.

It is enough to consider the case of twisting by $g \in H$, since any element of $N_G({\mathfrak{h}})$ is conjugated to an element from $H$. In this case subspaces of $\mathcal{H}_{\dot g}$ with all fixed fermion charges become representations of W-algebra~\footnote{This is a common well-known procedure, see e.g. \cite{MMMO} and references therein.}. The $r$-charges of the corresponding representations are given by shifts of the vector $\vec{r}=\frac{\log g}{2\pi i}$ by root lattice of~$\mathfrak{g}$.

Explicit formulas are collected below, but let us first discuss the irreducibility of representations. The Verma modules of W-algebras are irreducible if
\be
\label{eq:genericW}
\left(\alpha,r\right)\not\in\mathbb Z,
\ee
 see \cite{FKW}, \cite{Arakawa:2004} (in particular Theorem 6.6.1) or \cite{FL} (eq (4.4)). For generic $r$ this condition is satisfied and all modules, arising in the decomposition (subspaces of $\mathcal{H}_{\dot g}$ with all fixed fermion charges), are Verma modules due to coincidence of the characters.

If $g$ comes from the  element of $N_G({\mathfrak{h}})$ with nontrivial cyclic structure, then $r$ is not necessarily generic. For $\mathfrak{gl}(N)$ case, as follows \eqref{glsim}, the $r$-charges corresponding to a single cycle do satisfy \eqref{eq:genericW}, and for different cycles this condition also holds provided $r$ are generic (no relations between $r$ from different cycles). The same argument works for $\mathfrak{so}(2N)$ with ``plus-cycles'', but if we have at least two ``minus-cycles'', the corresponding $r$-charges can violate the condition \eqref{eq:genericW}, and not only Verma modules arise in the decomposition over irreducible representations.

In any case we have an identity of characters \eq{
\chi_{\dot g}(q)=\chi_0(q)\hat \chi_{\dot g}(q)
}
where $\chi_0(q)$ is the character of Verma module, and $\hat\chi_{\dot g}(q)$ is the character of the space of highest vectors. Hence, there is a non-trivial statement that all coefficients of the power expansion of the  ratios $\chi_{\dot g}(q)/\chi_0(q)$ are positive integers, which can be proven using identities, derived in the previous section.

The list of characters of the Verma modules, appeared above, is:
\begin{itemize}
\item $\mf{gl}(N)$, $\mf{so}(2N)$ (NS sector). Algebra is generated by $N$ bosonic currents, each of them producing $\frac1{\prod_{n>0}(1-q^n)}$, so the
character is
\eq{
\chi_0(q)=\frac1{\prod_{n=1}^\infty(1-q^n)^N}
}
\item $\mf{so}(2N)$ (R sector). One of these currents, $V(z)$, becomes Ramond, with half-integer modes:
\eq{
\chi_0(q)=\frac1{\prod_{n=1}^\infty(1-q^n)^{N-1}\prod_{n=0}^\infty(1-q^{\frac12+n})}
}
\item $\mf{so}(2N+1)$ (NS sector). One current, $V(z)$, becomes Neveu-Schwarz fermion, so taking into account its parity we get
\eq{
\chi^\pm_0(q)=\frac{\prod_{n=0}^\infty(1\pm q^{\frac12+n})}{\prod_{n=1}^\infty(1-q^n)^N}\label{Ver3}
}
\item $\mf{so}(2N+1)$ (R sector). In the case of the Ramond fermion $V(z)$ character $\chi_0^-(q)$ vanishes because the fermionic zero mode produces equal
numbers of states with the opposite fermionic parities:
\eq{
\chi^+_0(q)=2\frac{\prod_{n=1}^\infty(1+q^{n})}{\prod_{n=1}^\infty(1-q^n)^N}\\
\chi^-_0(q)=0
}
\end{itemize}

\paragraph{$\mf{gl}(N)$ case.}
Any element is conjugated to a product of cycles of length 1, so
\eq{
\hat\chi_{\dot g}(q)=q^{\Delta^0_g}\Sum_{\vec n\in Q_{A_{N-1}}} q^{\frac12(v+\vec n)^2}
}

\paragraph{$\mf so(2N)$ case, $K'=0$.}
Any element is conjugated to $\prod_{j=1}^N [1,e^{2\pi v_j}]_+$, so
\eq{
\hat\chi_{\dot g}(q)=q^{\Delta^0_g}\Sum_{\vec n\in Q_{D_{N}}} q^{\frac12(v+\vec n)^2}
}

\paragraph{$\mf so(2N)$ case, $K'>0$, NS-sector.}
Again, any element is conjugated to $\prod [1,e^{2\pi v_j}]_+$, so
\eq{
\hat\chi_{\dot g}(q)=2^{\frac{K'}2-1}q^{\Delta^0_g}\Sum_{\vec n\in Q_{B_{N}}} q^{\frac12(v+\vec n)^2}
}

\paragraph{$\mf{so}(2N)$ case, R-sector.}
Here any element is conjugated to $[1]_-\prod_{j=1}^{N-1} [1,e^{2\pi v_j}]_+$, so
\eq{
\hat\chi_{\dot g}(q)=2^{[\frac{K'}2]}q^{\Delta^0_g}\Sum_{\vec n\in Q_{B_{N-1}}} q^{\frac12(v+\vec n)^2}
}
because contribution from the cycle $[1]_-$ to the denominator cancels contribution from the Ramond boson $V(z)$.

\paragraph{$\mf{so}(2N+1)$ case, $K'=0$, NS fermion.}
Here one has two non-trivial characters
\eq{
\hat\chi^+_{\dot g}(q)=q^{\Delta^0_g}\Sum_{\vec n\in Q_{B_N}}q^{\frac12(v+\vec n)^2}\\
\hat\chi^-_{\dot g}(q)=q^{\Delta^0_g}\left(\Sum_{\vec n\in Q_{D_N}}q^{\frac12(v+\vec n)^2}-\Sum_{\vec n\in Q_{D_N'}}q^{\frac12(v+\vec n)^2}\right)
}

\paragraph{$\mf{so}(2N+1)$ case, $K'>0$} This case gives nothing interesting as compared to $D_N$ situation.
\eq{
\hat\chi^+_{\dot g}(q)=2^{[\frac{K'}2]}q^{\Delta^0_g}\Sum_{\vec n\in Q_{B_N}} q^{\frac12(v+\vec n)^2}\\
\hat\chi^-_{\dot g}(q)=0
}

\section{Characters from the lattice algebras constructions
\label{ss:lattice}}

\subsection{Twisted representation of $\widehat{\mathfrak{g}}_1$}

Now we reformulate the results of previous sections using the notion of \emph{twisted representations of vertex algebras}. Recall the corresponding setting (following, for example, \cite{Bakalov:2004}). Let $V$ be a vertex algebra (equivalently vacuum representation of the vertex algebra), and $\sigma$ be an automorphism of $V$ of finite order $l$. Then $V=\oplus V_k$, where $V_k=\{v\in V| \sigma v=\exp(2\pi ik/l)v\}$. A $\sigma$-twisted module is a vector-space $M$ endowed with a linear map
from $V$ to the space of currents
\be
v\mapsto A_v(z)=\sum_{m \in \frac1l\mathbb{Z}} a_m(v)z^{-m-1}, \quad v\in V,\; a_m(v) \in \mathrm{End}(M).
\ee
Such correspondence should be $\sigma$-equivariant, namely
\be
\label{bAv}
A_{\sigma v}(z)=A_v(e^{2\pi i}z)
\ee
giving the boundary conditions for the currents, and agree with the vacuum vector and relations in $V$. In particular, it follows from the $\sigma$-equivariancy \rf{bAv}, that if $v \in V_k$ then $A_v(z) \in z^{-k/l} \mathbb{C}[[z,z^{-1}]]$.

Consider now a Lie group $G$ (either $GL(N)$ or $SO(2N)$, $N \geq 2$), with $\mathfrak{g}={\rm Lie}(G)$ being the corresponding Lie algebra. Denote by $\mathrm{V}(\mathfrak{g})$ the irreducible vacuum representation of $\widehat{\mathfrak{g}}$ of the level one. This space has a structure of the vertex algebra, i.e. for any $v \in \mathrm{V}(\mathfrak{g})$ one can assign the current $A_v(z)$. This space of currents is generated by the currents $J_{\alpha \beta}(z)$ from Sect.~\ref{ss:WandKM}.

The vertex algebra $\mathrm{V}(\mathfrak{g})$ is a lattice vertex algebra. Let $Q_\mathfrak{g}$ denote the root lattice of $\mathfrak{g}$, and introduce rank of $\mathfrak{g}$ bosonic fields
with the OPE $\varphi_i(z)\varphi_j(w)=-\delta_{ij}\log(z-w)+\mathrm{reg}$, and the stress-energy tensor
$T(z)= - \half\sum_j :\partial \varphi_j(z)^2:$, then the currents of $\mathrm{V}(\mathfrak{g})$ can be presented in the bosonized form
\be
:\prod_{i,m} \partial^{a_{i,m}}\varphi_i \exp\left( i\sum_j\alpha_j\varphi_j(z)\right):,
\ee
where $\alpha=(\alpha_1,\ldots,\alpha_n)\in Q_{\mathfrak{g}}$ and $a_{i,m}$ are arbitrary positive integers, while the stress-energy tensor corresponding to standard conformal vector $\frac12 \sum J_{j,-1}^2|0\rangle = \tau \in \mathrm{V}(\mathfrak{g})$ (here $J_{j,n}$ are modes of the field $i\partial\varphi_j(z)$).
The group $G$ acts on $V(\mathfrak{g})$, and in order to use lattice algebra description we consider only the subgroup $\mr N_{G}(\mf h) \subset G$ which preserves the Cartan subalgebra.

In \cite{Bakalov:2004} the representations of the lattice vertex algebra, twisted by automorphisms, arise from isometries of the lattice $Q_\mathfrak{g}$. Here we restrict ourself to the isometries provided by action of the Weyl group $\texttt{W}$ (this case was actually considered in \cite{Kac:112} without language of twisted representations).
Let $s\in \texttt{W}$ be an element of the Weyl group, by $g$ we denote its lifting to $G$, in other words $g \in \mr N_{G}(\mf h)$ such that adjoint action $g$ on $\mathfrak{h}$ coincides with $s$. We consider representation twisted by such $g$. Setting of \cite{Bakalov:2004} and \cite{Kac:112} works for special $g$, for example such $g$ should have finite order, but we will expand this to the generic
$g \in \mr N_{G}(\mf h)$. Clearly, the conformal vector $\tau$ is invariant under the adjoint action of $N_{G}(\mf h)$.

The $g$-twisted representations of $V(\mathfrak{g})$ in \cite{Bakalov:2004} are defined as  a direct sum of twisted representations of $\widehat{\mathfrak{h}}$. By $\{e^{2\pi i\theta_{\mathrm{Adj},k}}\}$ we denote eigenvalues of $s$, or of the adjoint action $g_{\rm adj}$ on $\mathfrak{h}$, we set $-1< \theta_{\mathrm{Adj},k}\leq 0$, by $\{J_k\in \mathfrak{h}\}$ --- the corresponding eigenvectors, and define the currents
\be
J_k(z)=\sum_{n \in \mathbb{Z}}J_{k,\theta_{\mathrm{Adj},k}+n}z^{-\theta_{\mathrm{Adj},k}-n-1}
\ee
A $g$-twisted representation of the Heisenberg algebra $\widehat{\mathfrak{h}}$ is a Fock module $\mathrm{F}_\mu$ with the highest weight vector $v_\mu$
\be
\label{JFock}
J_{k,\theta_{\mathrm{Adj},k}+n}v_\mu =0, n> 0,\qquad J_{k,0}v_\mu =\mu(J_k)v_\mu,\; \text{if}\ \ \  \theta_{\mathrm{Adj},k}=0.
\ee
generated by creation operators $J_{k,\theta_{\mathrm{Adj},k}+n}$, $n \leq 0$. Here $\mu \in \mathfrak{h}_0^*$, where $\mathfrak{h}_0$ is
$g_{\rm adj}$-invariant subspace of $\mathfrak{h}$.

It has been proven in \cite{Bakalov:2004} that twisted representations of $V(\mathfrak{g})$ have the structure
\be
\label{eq:BKtwisted}
M(s,\mu_0)=\oplus_{\mu\in \mu_0+\pi_s Q_\mathfrak{g}}\mathrm{F}_\mu\otimes \mathbb{C}^{d(s)}
\ee
for certain finite set of $\mu_0 \in \mathfrak{h}_0^*$. Here $\pi_s$ denotes projection from $\mathfrak{h}^*$ to $\mathfrak{h}_0^*$, corresponding to the element $s\in \texttt{W}$ for the chosen adjoint action $g_{\rm adj}$. For any root $\alpha$ the corresponding current $J_\alpha(z)$ acts from $\mathrm{F}_\mu$ to $\mathrm{F}_{\mu+\pi_s\alpha}$ and equals to the linear combination of vertex operators.
Number $d(s)$ denotes the defect of the element $s \in \texttt{W}$, its square is defined by
\be d(s)^2=|(Q_{\mathfrak{g}}\cap \mathfrak{h}_0^\perp)/(1-s)P_{\mathfrak{g}}|.
\ee
Here $P_{\mathfrak{g}}$ denotes the weight lattice of $\mathfrak{g}$, $\mathfrak{h}_0^\perp$ denotes the space of linear functions vanishing on $\mathfrak{h}_0$, $|\cdot|$ stands for the number of elements in the group. It can be proven that for any $s$ the number $d(s)$ is integer. In our case ($GL(N)$ and $SO(n)$ groups) this number always equals to some power of $2$.

Formula \eqref{eq:BKtwisted} allows to calculate the character of the module $M$, i.e. the trace of $q^{L_0}$. First, notice that the character of the Fock module $\mathrm{F}_\mu$ equals
\be
\chi_\mu(q) = \frac{q^{\Delta_\mu}}{\prod_i\prod_{n=1}^\infty(1-q^{\theta_{\mathrm{Adj},i}+n})}
\ee
 where $\Delta_\mu$ is an eigenvalue of $L_0$ on the vector $v_\mu$. The value of $\Delta_\mu$ consists of two contributions. The first comes from the terms with $\theta_{\mathrm{Adj}}=0$, and, as follows from \rf{JFock}, is equal to
$\half(\mu,\mu)$. The second contribution comes from the normal ordering. The vectors $J_k\in \mathfrak{h}$, corresponding to $\theta_{\mathrm{Adj},k} \neq 0$ can always be arranged into orthogonal pairs $(J_1, J_{1'})$, $(J_2, J_{2'})$, \ldots with complementary eigenvalues $\theta_{\mathrm{Adj},k}+\theta_{\mathrm{Adj},k'}=-1$~\footnote{There is also ``degenerate'' case $J_k=J_{k'}$ for $\theta_{\mathrm{Adj},k}=\theta_{\mathrm{Adj},k'}=-\half$.}. After normal ordering of the corresponding currents one gets
\be
J_k(z)J_{k'}(w)=\sum_{n,m\in \mathbb{Z}} \frac{J_{k,n+\theta}}{z^{n+\theta+1}}\frac{J_{k',m-\theta}}{w^{m-\theta+1}}=
\sum_{n\in\mathbb{Z},m\geq 0} \frac{J_{k,n+\theta}}{z^{n+\theta+1}}\frac{J_{k',m-\theta}}{w^{m-\theta+1}}+
\\
+\sum_{n\in\mathbb{Z},m< 0} \frac{J_{k',m-\theta}}{w^{m-\theta+1}}\frac{J_{k,n+\theta}}{z^{n+\theta+1}}+
\sum_{n>0} (n+\theta)\frac{w^{n+\theta-1}}{z^{n+\theta+1}}
\ee
where  $\theta=\theta_{\mathrm{Adj},k}$. The last term in the r.h.s., which appears due to $\left[J_{k,n+\theta},J_{k',m-\theta}\right]=(n+\theta)\delta_{n+m,0}$\,, also gives a nontrivial contribution to the action of $L_0$ on highest vector $v_\mu$, since
\be
\sum_{n>0} (n+\theta)\frac{w^{n+\theta-1}}{z^{n+\theta+1}} = \frac{(1+\theta)w^\theta z^{-\theta}+(-\theta)w^{1+\theta}z^{-1-\theta}}{(z-w)^2}=
\\
\stackreb{z\to w}{=}\ \frac{1}{(z-w)^2}-\frac{\theta(1+\theta)}{2w^2}+\mathrm{reg}
\ee
Altogether one gets
\be
\Delta_\mu=\frac12(\mu,\mu)-\sum_k\frac{\theta_{\mathrm{Adj},k}(1+\theta_{\mathrm{Adj},k})}4
\ee
and therefore, finally for the character of \rf{eq:BKtwisted}

\be
\label{eq:chgen}
\mathrm{Tr}q^{L_0}\Bigr|_{M(s,\mu_0)}=q^{-\frac14\sum_k\theta_{\mathrm{Adj},k}(1+\theta_{\mathrm{Adj},k})}
\frac{d(s)\sum_{\mu\in \mu_0+\pi_w Q}q^{\frac12(\mu,\mu)}}{\prod_{i=1}^{N}\prod_{n=1}^\infty(1-q^{\theta_{\mathrm{Adj},i}+n})}
\ee

Recall that the initial weight $\mu_0$ in the setting of \cite{Bakalov:2004} should belong to the finite set in $\mathfrak{h}_0^*$ (or $\mathfrak{h}_0^*/\pi_{W}Q$). But we will generalize such representations and take any $\mu_0\in\mathfrak{h}_0^*$. This can be viewed as a twisting by more general elements $g \in \mr N_{G}(\mf h)$, which can have infinite order. Actually the corresponding elements are representatives of the conjugacy classes of $\mr N_{G}(\mf h)$ used in Sect.~\ref{ss:twists}.

\subsection{Calculation of characters}

\paragraph{$GL(N)$ case}
The root lattice $Q_{\mathfrak{gl}(N)}=Q_{{A}_{N-1}}$ is generated by the vectors $\{e_i-e_j\}$, where  $\{e_1,\ldots,e_N\}$ denote the vectors of orthonormal basis in $\mathbb{R}^N$. Assume that $s\in \texttt{W}$ is a product of disjoint cycles of lengths $l_1,\ldots,l_K$, then without loss of generality the action of such elements can be defined as $(e_1\mapsto e_2\mapsto\ldots \mapsto e_{l_1} \mapsto e_1), (e_{l_1+1}\mapsto e_{l_1+2}\mapsto\ldots \mapsto e_{l_1+l_2} \mapsto e_{l_1+1}),\ldots $.

In this case $\mathfrak{h}_0^*$ (the $s$-invariant part of $\mathfrak{h}^*$) is generated by the vectors
\be
\label{fe}
f_1=e_1+\ldots+e_{l_1},\ \  f_2=e_{l_1+1}+\ldots+e_{l_1+l_2},\ \  \ldots
\ee
 while
$\pi_s Q_{\mathfrak{gl}(N)}$ is generated by the vectors $\frac1{l_i}f_i-\frac1{l_j}f_j$, so one can present any element of $\pi_s Q_{\mathfrak{gl}(N)}$ as $\sum \frac1{l_j}n_jf_j$ with $\sum n_j=0$ and identify with that from $Q_{\mathfrak{gl}(K)}$. Let $\mu_0=\sum_{j} r_jf_j$. Then the formula \eqref{eq:chgen} takes here the form
\be
\mathrm{Tr}(q^{L_0})\bigr|_{M(s,\mu_0)}=q^{\Delta_s^0}\frac{\sum_{Q_{\mathfrak{gl}(K)}}  q^{\sum_j\frac1{2l_j}(n_j+l_jr_j)^2}}{\prod_{j=1}^{K}\prod_{n=1}^\infty(1-q^{n/l_j})},
\label{glNtw}
\ee
where, since for any length $l$ cycle $\theta_{\mathrm{Adj},k}=-k/l$,
\be \label{eq:Deltas1}
\Delta^0_s=\sum_{j=1}^K\sum_{i=1}^{l_j}\frac{i(l_j-i)}{4l_j^2} = \sum_{j=1}^K \frac{l_j^2-1}{24l_j}
\ee
This formula coincides with \eqref{eq:chigln}, and the reason is that the corresponding element from $\mr N_{GL(N)}(\mf h)$ is exactly \rf{glNg}, $g=\prod_{j=1}^K [l_j,e^{2\pi ir_j}]$. Indeed, let $\alpha=e_a-e_b$, where $a$ belongs to the cycle $j$ and $b$  belongs to the cycle $j'$\,, then the current $J_\alpha(z)$ shifts $L_0$ grading by $r_j-r_{j'}+\text{[rational number with denominator $l_j l_{j'}]$}$.

\paragraph{$SO(2N)$ case} The root lattice $Q_{\mathfrak{so}(2N)}=Q_{D_N}$ is generated by the vectors $\{e_i-e_j, e_i+e_j\}$, where again $e_1,\ldots,e_N$ denote the basis in $\mathbb{R}^N$. As we already discussed in Sect.~\ref{ss:twists}, there are two types of the Weyl group elements, the first type just permutes $e_i$, while the second type permutes $e_i$ together with the sign changes.

The first case almost repeats the previous paragraph, without loss of generality we assume that the Weyl group element acts as $(e_1\mapsto e_2\mapsto\ldots \mapsto e_{l_1} \mapsto e_1), (e_{l_1+1}\mapsto e_{l_1+2}\mapsto\ldots \mapsto e_{l_1+l_2} \mapsto e_{l_1+1}),\ldots $, where $l_1,\ldots,l_K$ are again the lengths of the cycles. The $s$-invariant part of $\mathfrak{h}^*_0$ is generated by the same ``averaged'' vectors \rf{fe}, while $\pi_sQ_{D_N}$ is generated by the vectors $\frac1{l_i}f_i-\frac1{l_j}f_j, \frac1{l_i}f_i+\frac1{l_j}f_j$. In other words, $\pi_sQ_{D_N}$ consist of the vectors $\sum \frac{n_j}{l_j}f_j$, where $(n_1,\ldots,n_k)\in Q_{\mathfrak{so}(2K)}$. Let $\mu_0=\sum_{j} r_jf_j$, then the character formula \rf{eq:chgen} for this case acquires the form
\be
\mathrm{Tr}(q^{L_0})\bigr|_{M(s,\mu_0)}=q^{\Delta_s^0}\frac{\sum_{Q_{\mathfrak{so}(2K)}} q^{\sum_j\frac1{2l_j}(n_j+l_jr_j)^2}}{\prod_{j=1}^{K}\prod_{n=1}^\infty(1-q^{n/l_j})}
\ee
and coincides with \eqref{chiD1}. Here $\Delta_s^0$ is defined in \eqref{eq:Deltas1}. The corresponding element from $\mr N_{SO(2N)}(\mf h)$ has the form $\prod_{j=1}^K[l_j,e^{2\pi i r_j}]_+$ in the notations of Sect.~\ref{ss:twists} (see \rf{olemma}).

For the second type (the corresponding element from $\mr N_{SO(2N)}(\mf h)$ has the form $\prod_{j=1}^K[l_j,e^{2\pi i r_j}]_+\cdot\prod_{j'=1}^{K'}[l_{j'}]_-$) one can present the Weyl group element as a product of $K$ disjoint cycles of lengths $l_1,\ldots,l_K$ which just permute $e_i$, and $K'$ cycles of lengths $l_{1'},\ldots,l_{K'}$ which permute $e_i$ with signs, see \rf{olemma}.  Now, without loss of generality, we assume that $s$ acts as $(e_1\mapsto e_2\mapsto\ldots \mapsto e_{l_1} \mapsto e_1), (e_{l_1+1}\mapsto e_{l_1+2}\mapsto\ldots \mapsto e_{l_1+l_2} \mapsto e_{l_1+1}),\ldots $, $(e_{L+1}\mapsto e_2\mapsto\ldots \mapsto e_{L+l_{1'}} \mapsto -e_1 ), (e_{L+l_{1'}+1}\mapsto e_{L+l_{1'}+2}\mapsto\ldots \mapsto e_{L+l_{1'}+l_{2'}} \mapsto -e_{L+l_{1'}+1}),\ldots $, where $L=l_1+\ldots+l_K$.  The $s$-invariant part of $\mathfrak{h}^*_0$ is generated by the same vectors \rf{fe}, while $\pi_sQ_{D_N}$ is generated by the vectors $\frac1{l_i}f_i$. One can say that $\pi_sQ_{D_N}$ consists of the vectors $\sum  \frac{n_j}{l_j}f_j$, where $(n_1,\ldots,n_k)\in Q_{\mathfrak{so}(2K+1)}=Q_{B_N}$, so that for the character formula one gets
\be	
\mathrm{Tr}(q^{L_0})\bigr|_{M(s,\mu_0)}=q^{\Delta_s^0}\frac{2^{K'/2-1}\sum_{Q_{\mathfrak{so}(2K+1)}} q^{\sum_j\frac1{2l_j}(n_j+l_jr_j)^2}}{\prod_{j=1}^{K}
\prod_{n=1}^\infty(1-q^{n/l_j})\prod_{j=1}^{K'}\prod_{n=1}^\infty(1-q^{(2n-1)/2l'_j})},
\ee
where, since in addition to $[l]_+$-cycles with $\theta_{\mathrm{Adj},k}=-k/l$ one has now $[l']_-$-cycles
with $\theta_{\mathrm{Adj},k}'=-(k-\half)/l'$,
\be
\Delta_s^0=\sum_{j=1}^K\sum_{i=1}^{l_j}\frac{i(l_j-i)}{4l_j^2}
+\sum_{j=1}^{K'}\sum_{i=1}^{l'_j}\frac{(2i-1)(2l'_{j}-2i+1)}{16l'^2_{j}} =
\\
= \sum_{j=1}^K \frac{l_j^2-1}{24l_j} +\sum_{j=1}^{K'} \frac{2l'^2_{j}+1}{48 l'_{j}}
\label{dimD0}
\ee
This formula coincides with \eqref{chiD2}. The number $2^{K'/2-1}$ equals to $d(\sigma)$, this is the first case where this number is nontrivial. Note, that we consider here only internal automorphisms, i.e. $K'$ is even.

Recall also (see Sect.~\ref{ss:Chariden}) that if $g,g' \in \mr N_{G}(\mf h)$ are conjugate in $G$\,, then corresponding characters $\mathrm{Tr}(q^{L_0})\bigr|_{M(s,\mu_0)}$ and $\mathrm{Tr}(q^{L_0})\bigr|_{M(s',\mu'_0)}$ are equal.

\subsection{Characters from principal specialization of the Weyl-Kac formula}

Fix an element $g \in G$ of finite order $l$. The $g$-twisted representations of $V(\mathfrak{g})$ are representations of the affine Lie algebra twisted by $g$. Recall that these twisted affine Lie algebras $\widehat{\mathcal{L}}(\mathfrak{g},g)$ are defined in \cite[Sec. 8]{KacBook} as $g$ invariant part of $\mathfrak{g}[t,t^{-1}]\oplus \mathbb{C}\texttt{k}$, where $g$ acts as
\begin{equation}
g(t^j\otimes J)=\epsilon^{-j}t^j\otimes \left(gJg^{-1}\right), \text{ where } \epsilon=\exp(2\pi i/l),\quad g(\texttt{k})=\texttt{k}.
\end{equation}
By definition $g$ is an internal automorphism, therefore the algebra $\widehat{\mathcal{L}}(\mathfrak{g},g)$ is isomorphic to $\widehat{\mathfrak{g}}$ (see Theorem \cite[8.5]{KacBook}), though natural homogeneous grading on $\widehat{\mathcal{L}}(\mathfrak{g},g)$ differs from the homogeneous grading on $\widehat{\mathfrak{g}}$.

Therefore the  $g$-twisted representations of $V(\mathfrak{g})$ as a vector spaces are integrable representations of $\widehat{\mathfrak{g}}$ \footnote{Note that we get only level 1 integrable representation of $\widehat{\mathfrak{g}}$, since  $V(\mathfrak{g})$ was defined above as a lattice vertex algebra, i.e. vacuum representation of the level $\texttt{k}=1$}.
Their characters can be computed using the Weyl-Kac character formula.
This formula has simplest form in the principal specialization, i.e. computed on the element $q^{\rho^\vee}\in\widehat{G}$. Here $\rho^\vee\in \mathfrak{h}\oplus \mathbb{C}\texttt{k}$ is such that $\alpha_i(\rho^\vee)=1$ for all affine simple roots $\alpha_i$ (including $\alpha_0$). Then the character of integrable highest weight module with the highest weight $\Lambda$ equals (see \cite[eq. (10.9.4)]{KacBook})
\be
\mathrm{Tr}(q^{\rho^\vee/h})\bigr|_{\mathrm{L}_\Lambda}=q^{\Lambda(\rho^\vee)/h}\prod_{\alpha^\vee \in \Delta^\vee_+}\left(\frac{1-q^{(\Lambda+\rho,\alpha^\vee)/h}}
{1-q^{(\rho,\alpha^\vee)/h}}\right)^{\operatorname{mult}(\alpha^\vee)},
\label{eq:princpSpec}
\ee
where $\Delta^\vee_+$ is the set of all positive (affine) coroots.
Here $h$ is the Coxeter number. It will be convenient to use $q^{\rho^\vee/h}$ instead of $q^{\rho^\vee}$. The weight $\rho$ is defined by $(\rho,\alpha_i^\vee)=1$ for all simple coroots $\alpha_i$ (including affine root $\alpha_0$).

The grading above in this section was the $L_0$ grading, and it was obtained using the twist by the element $g\in \mathrm{N}_G(\mathfrak{h})$. Now we take certain $g$ such that $g$-twisted $L_0$ grading coincides with principal grading in \eqref{eq:princpSpec}. We take $g$ in the Cartan subgroup $H$ and, as was explained above, choice of $g$ corresponds to the choice of $\mu_0$ in \eqref{eq:chgen}.

In the principal grading used in \eqref{eq:princpSpec} $\deg E_{\alpha_i}=\frac1h$ for all simple roots $E_{\alpha_i}$  (including affine root $\alpha_0$). Therefore $\mu_0 \in
P_\mathfrak{g}+\frac1{h}\overline{\rho}$, where $P_\mathfrak{g}$ is the weight lattice for $\mathfrak{g}$, $\overline{\rho}$ is defined by the formula $(\overline{\rho},\alpha_i)=1$ for all simple roots\footnote{Note the difference between $\rho$ and $\overline{\rho}$: the first was defined by pairing with simple coroots (including affine one), and the second is defined by scalar products with (non affine) roots. In the simply laced case these conditions in terms of roots and coroots are  equivalent and we have $\rho=\overline{\rho}+h\Lambda_0$}.

Below we write explicit formulas for the characters of twisted representation corresponding to such $g$ (and such $\mu$). In the simply laced case, computing the characters using two formulas \eqref{eq:chgen} and \eqref{eq:princpSpec} one gets an identity, which is actually the Macdonald identity \cite{Macdonald:1972}.

In notations for the root system we follow \cite{Macdonald:1972} and \cite{KacBook}. Below we consider roots as vectors in the linear space, generated by $e_1,\ldots,e_n,\delta,\Lambda_0$, and coroots --- in the space generated by $e_1^\vee,\ldots,e_n^\vee,K,d$. The pairing between these dual spaces given by $(e_i,e_j^\vee)=\delta_{ij}$, $(\Lambda_0,K)=(\delta,d)=1$, while all other vanish.

\paragraph{$GL(N)$ case.} Root system is $A_{N-1}^{(1)}$ (affine $A_{N-1}$), dual root system is also $A_{N-1}^{(1)}$.
\begin{equation}
\begin{aligned}
&\text{Simple roots: }\ \ \ \ \ \  \alpha_0=\delta-e_1+e_N,\;\; \alpha_i=e_i-e_{i+1}, \, 1\leq i \leq N-1\\
&\text{Simple coroots: } \alpha^\vee_0=K+e_N^\vee-e_1^\vee,\;\; \alpha^\vee_i=e_{i}^\vee-e_{i+1}^\vee,\, 1\leq i\leq N-1\\
&\text{Real coroots: }\ \ \ \ \ \ \ \ \ \ \  mK+e_i^\vee-e^\vee_j,\;\;  m \in \mathbb{Z}, i \neq j\\
&\text{Imaginary coroots: }\ \ \  mK \text{ of multiplicity } N,\; m\in \mathbb{Z}.\\
&\text{Level $\texttt{k}=1$ weights: }\ \  \Lambda_0,\;\; \Lambda_j=\Lambda_0+\sum_{i=1}^{j}e_i,  1\leq j \leq N-1\\
&h=N,\; \rho=\half\sum_{i=1}^N (N-2i+1)e_i+N\Lambda_0,\; \overline{\rho}=\half\sum_{i=1}^N (N-2i+1)e_i.
\end{aligned}
\label{glNdefs}
\end{equation}
Note that the multiplicity of imaginary roots is $N$ instead on $N-1$ since we consider $G=GL(N)$ instead of $SL(N)$, and the corresponding affine algebra differs by one additional Heisenberg algebra.

The computation of the denominator in \eqref{eq:princpSpec}, using \rf{glNdefs} gives
\be
\prod_{\alpha^\vee \in \Delta^\vee_+}
(1-q^{(\rho,\alpha^\vee)/h})^{\operatorname{mult}(\alpha^\vee)}=\prod_{k=1}^{\infty}(1-q^{k/N})^{N}
\ee
while for the numerator (the same for all level $\texttt{k}=1$ weights) one gets
\be
\prod_{\alpha^\vee \in \Delta^\vee_+}
(1-q^{(\Lambda+\rho,\alpha^\vee)/h})^{\operatorname{mult}(\alpha^\vee)}=\prod_{k=1}^{\infty}(1-q^{k/N})^{N-1}
\ee
so that the character \eqref{eq:princpSpec} in principal specialization
\be
q^{-\Lambda(\rho^\vee)/h}\mathrm{Tr}(q^{\rho^\vee/h})\bigr|_{\mathrm{L}_\Lambda}=\frac{1}{\prod_{k=1}^\infty (1-q^{k/N})}
\ee
One can compare the last expression with the formula \eqref{eq:chgen} using the choice of $\mu_0$, as explained above. We get an identity
\be
\label{eq:LW:chiAn}
\frac{\Sum_{\alpha \in Q_{\mathfrak{sl}(N)}} q^{\frac12(\alpha+\frac1N\overline{\rho},\alpha+\frac1N\overline{\rho})}}{\prod_{k=1}^\infty(1-q^k)^N}
=\frac{q^{\frac1{2N^2}(\overline{\rho},\overline{\rho})}}{\prod_{k=1}^\infty (1-q^{k/N})}.
\ee
which is a particular case of formula \rf{glNidt} from Sect.~\ref{ss:Chariden}, and again reproduces the product formula for the lattice $A_{N-1}$ theta function \rf{ANprod}.

Recall that the r.h.s. of \eqref{eq:LW:chiAn} also has an interpretation of a character of the twisted Heisenberg algebra. This twist of the Heisenberg algebra emerges in the representation twisted by $g$ with $g_{\mathrm{ Adj}}$ acting as the Coxeter element of the Weyl group, hence  r.h.s. of \eqref{eq:LW:chiAn}
equals to the r.h.s. of \rf{glNtw} for a single cycle $K=1$, $l=N$.
This $g$ is conjugate to used above in the computation of l.h.s., therefore the characters of the twisted modules should be the same.
The construction of level one representations in terms of principal Heisenberg subalgebra is
well-known, see \cite{Lepowsky-Wilson, Kac-Lepowsky_Wilson}.
Another interpretation of the l.h.s in \eqref{eq:LW:chiAn} is the sum of characters of the W-algebra, namely W-algebra of $\mathfrak{gl}(N)$, (see Sect.~\ref{ssec:W:decom}).

\paragraph{$SO(2N)$ case.} Root system $D_{N}^{(1)}$ (affine $D_{N}$), the dual root system is also $D_{N}^{(1)}$.
\begin{equation}
\begin{aligned}
&\text{Simple roots: } \alpha_0=\delta{-}e_1{-}e_2,\;\; \alpha_i=e_i{-}e_{i+1}, \, 1\leq i <N, \alpha_N=e_{N-1}{+}e_N\\
&\text{Simple coroots: } \alpha^\vee_0=K-e_1^\vee{-}e_2^\vee,\; \alpha^\vee_i=e_{i}^\vee{-}e_{i+1}^\vee,\, 1\leq i<N,\; \alpha_N=e^\vee_{N-1}{+}e^\vee_N\\
&\text{Real coroots: } mK{+}e_i^\vee{-}e^\vee_j,\; mK+e^\vee_i+e^\vee_j,\; mK{-}e^\vee_i{-}e^\vee_j, m \in \mathbb{Z}, i \neq j\\
&\text{Imaginary coroots: } mK \text{ of multiplicity } N,\; m\in \mathbb{Z}\\
&\text{$\texttt{k}=1$ weights: } \Lambda_0,\; \Lambda_1{=}e_1{+}\Lambda_0,  \;\Lambda_{N-1}{=}\half\sum_{i=1}^Ne_i{+}\Lambda_0,\; \Lambda_{N}{=}\half\sum_{i=1}^Ne_i{-}e_N{+}\Lambda_0 \\
&h=2N-2,\;\; \rho=\sum_{i=1}^N (N-i)e_i+(2N-2)\Lambda_0,\;\overline{\rho}=\sum_{i=1}^N (N-i)e_i.
\end{aligned}
\end{equation}
Now again we just compute the denominator
\be
\prod_{\alpha^\vee \in \Delta^\vee_+}
(1-q^{(\Lambda+\rho,\alpha^\vee)/h})^{\operatorname{mult}(\alpha^\vee)}=\prod_{k=1}^{\infty}(1-q^{k/(2N-2)})^{N}
\ee
and the numerator (the same for all $\texttt{k}=1$ weights)
\be
\prod_{\alpha^\vee \in \Delta^\vee_+}
(1-q^{\frac{(\rho,\alpha^\vee)}{2N-2}})^{\operatorname{mult}(\alpha^\vee)}=
\\
= \prod_{j=1}^{N-1}\prod_{k=1}^{\infty}(1-q^{k-\frac{2j-1}{2N-2}})^{N+1}(1-q^{k-\frac{2j}{2N-2}})^{N}\cdot \prod_{k=1}^{\infty}(1-q^{k-\frac12})
\ee
in \eqref{eq:princpSpec}, giving for the character
\be
q^{-\Lambda(\rho^\vee)/h}\mathrm{Tr}(q^{\rho^\vee/h})\bigr|_{\mathrm{L}_\Lambda}=\prod_{j=1}^{N-1}\prod_{k=1}^\infty (1-q^{k-\frac{2j-1}{2N-2}})^{-1}\cdot \prod_{k=1}^\infty (1-q^{k-\frac{1}2})^{-1}.
\ee
As in the previous case, comparing this with the formula \eqref{eq:BKtwisted}, one gets an identity
\be
\label{eq:LW:chiDn}
\frac{\sum_{\alpha \in Q_{D_N}} q^{\frac12(\alpha+\frac1h\overline{\rho},\alpha+\frac1h\overline{\rho})}}{\prod_{k=1}^\infty(1-q^k)^N}=
\frac{q^{\frac1{2h^2}(\overline{\rho},\overline{\rho})}}{\prod_{j=1}^{N-1}\prod_{k=1}^\infty (1-q^{k-\frac{2j-1}{2N-2}})\cdot \prod_{k=1}^\infty (1-q^{k-\frac{1}2})}.
\ee
where the r.h.s. can be interpreted as a character of the representation of Heisenberg algebra twisted by $g$, such that $g_{\mathrm{Adj}}$ is Coxeter element. Again, this is the same as construction of the level $\texttt{k}=1$ representation in terms of principal Heisenberg subalgebra from
\cite{Lepowsky-Wilson,Kac-Lepowsky_Wilson}.
The l.h.s. of the formula \eqref{eq:LW:chiDn} can also be interpreted as the sum of characters of the $W(\mathfrak{so}({2N}))$-algebra, (see Sect.~\ref{ssec:W:decom}).

By now in this section we have considered only the simply laced case --- the only one, when the algebra $V(\mathfrak{g})$ is the lattice algebra or, in other words, when the level $\texttt{k}=1$ representations can be constructed as a sum of representations of the Heisenberg algebra. However, the formula \eqref{eq:princpSpec} is valid for any affine Kac-Moody algebra. Below we consider the case $G=SO(2N+1)$, where the level $\texttt{k}=1$ representations can be constructed using free fermions.

\paragraph{$SO(2N+1)$, $N>2$ case.} Root system is  $B_N^{(1)}$ (affine $B_N$), the dual root system  is $B_N^{(1,\vee)}=A_{2N-1}^{(2)}$ (affine twisted $A_{2N-1}$)
\begin{equation}
\begin{aligned}
&\text{Simple roots: } \alpha_0=\delta{-}e_1{-}e_2,\;\; \alpha_i=e_i{-}e_{i+1},\, 1\leq i \leq N{-}1,\;\; \alpha_N=e_N.\\
&\text{Simple coroots: }\alpha_0^\vee=K-e_1^\vee{-}e_2^\vee,\;\; \alpha^\vee_i=e_i^\vee{-}e_{i+1}^\vee,\,1\leq i \leq N{-}1,\;\; \alpha_N=2e_N^\vee.\\
&\text{Real coroots: } 2m K{\pm} 2e_i,\, mK{\pm} e_i{\mp} e_j,\,m K{\pm} e_i{\pm} e_j,\, 1\leq i <j \leq N,\, m\in \mathbb{Z}.\\
&\text{Imaginary coroots: } (2m-1)K\, \text{of multiplicity $N-1$},\, m\in \mathbb{Z}\, \\
&\qquad \qquad \qquad\qquad 2mK\, \text{of multiplicity $N$}, m\in \mathbb{Z}\setminus\{0\}.\\
&\text{$\texttt{k}=1$ weights: } \Lambda_0,\; \Lambda_1=\Lambda_0 + e_1,\;\; \Lambda_N=\Lambda_0+\half\sum\nolimits_{i=1}^N e_i \\
&h=2N,\;\rho=\sum\nolimits_{j=1}^N (N-j+\half)e_j+(2N-1)\Lambda_0,\; \;\;\overline{\rho}=\sum\nolimits_{j=1}^N (N-j+1)e_j.
\end{aligned}
\end{equation}
Compute again the denominator
\be
\prod_{\alpha^\vee \in \Delta^\vee_+}\left(1-q^{\frac{(\rho,\alpha^\vee)}{2N}}\right)^{\operatorname{mult}(\alpha^\vee)}
=\prod_{k=1}^\infty (1-q^{\frac{k}{2N}})^{N}
\cdot \prod_{k=1}^\infty (1-q^{\frac{2k-1}{2N}})
\label{eq:BNden}
\ee
and the numerator in the formula \eqref{eq:princpSpec}. Now the numerator for $\Lambda=\Lambda_0$ and $\Lambda=\Lambda_1$ is the same 	
\be
\prod_{\alpha^\vee \in \Delta^\vee_+}\left(1-q^{\frac{(\rho+\Lambda_0,\alpha^\vee)}{2N}}\right)=\prod_{\alpha^\vee \in \Delta^\vee_+}\left(1-q^{\frac{(\rho+\Lambda_1,\alpha^\vee)}{2N}}\right)=
\\
= \prod_{k=1}^\infty (1-q^{\frac{k}{2N}})^{N} \cdot \prod_{k=1}^\infty (1+q^{k})
\label{eq:BNnum1}
\ee
but for $\Lambda=\Lambda_N$ it is different
\be
\prod_{\alpha^\vee \in \Delta^\vee_+} \left(1-q^{\frac{(\rho+\Lambda_N,\alpha^\vee)}{2N}}\right)=\prod_{k=1}^\infty (1-q^{\frac{k}{2N}})^{N} \cdot \prod_{k=1}^\infty (1+q^{k-\frac12}), \label{eq:BNnum2}	
\ee
Here we used the identities \rf{simpid} and $\prod_{k=1}^\infty (1-q^{2k-1})(1-q^{k-1/2})^{-1}=\prod_{k=1}^{\infty}(1+q^{k-1/2})$. It is convenient to consider the direct sums of two representations $\mathrm{L}_{\Lambda_0}\oplus \mathrm{L}_{\Lambda_1}$ and $\mathrm{L}_{\Lambda_N}\oplus \mathrm{L}_{\Lambda_N}$ since these sums have construction in terms of fermions. Using \eqref{eq:princpSpec} one gets
\be
q^{-\Lambda_0(\rho^\vee)/h}\mathrm{Tr}(q^{\rho^\vee/h})\bigr|_{\mathrm{L}_{\Lambda_0}}+q^{-\Lambda_1(\rho^\vee)/h}\mathrm{Tr}(q^{\rho^\vee/h})\bigr|_{\mathrm{L}_{\Lambda_1}}=2\prod_{k=1}^\infty \frac{(1+q^{k})}{(1-q^{\frac{2k-1}{2N}})},\\
q^{-\Lambda_N(\rho^\vee)/h}\mathrm{Tr}(q^{\rho^\vee/h})\bigr|_{\mathrm{L}_{\Lambda_N}}=2\prod_{k=1}^\infty \frac{(1+q^{k-\frac12})}{(1-q^{\frac{2k-1}{2N}})}.
\ee
The r.h.s. of these equations suggests the existence of the construction of these representations in terms of $N$-component twisted (principal) Heisenberg algebra and additional fermion (in NS and R sectors correspondingly), exactly this construction has been considered in Sect.~\ref{ssec:bos:so(n)}.

On the other hand these characters can be rewritten in terms of the simplest $B$-lattice theta-functions just using the Jacobi triple product identity
\be
2\prod_{k=1}^\infty \frac{(1+q^{k})}{(1-q^{\frac{2k-1}{2N}})}=\prod_{k=1}^\infty \prod_{i=0}^{2N} (1+q^{k-\frac{i}{2N}})=
\\
= \sum_{n_1,\ldots,n_N \in \mathbb{Z}} q^{\frac12\sum_{j=1}^{N} (n_j^2+\frac{j}{N} n_j)} \prod_{k=1}^\infty\frac{(1+q^{k-\frac12})}{(1-q^{k})^N}
=\\= q^{-\frac{(N+1)(2N+1)}{48N}}\sum_{\alpha \in Q_{B_N}}q^{\frac12 (\alpha+\frac1{2N}\overline{\rho},\alpha+\frac1{2N}\overline{\rho})}
\prod_{k=1}^\infty\frac{(1+q^{k-\frac12})}{(1-q^{k})^N},
\ee
and	
\be
2\prod_{k=1}^\infty \frac{(1+q^{k-\frac12})}{(1-q^{\frac{2k-1}{2N}})}= 2\prod_{k=1}^\infty \prod_{i=0}^{2N-1} (1+q^{k-\frac{i}{2N}}) \prod_{k=1}^\infty (1+q^{k-\frac12})=
\\
= \sum_{n_1,\ldots,n_N \in \mathbb{Z}} q^{\frac12\sum_{j=1}^N (n_j^2+\frac{(j-1)}{N} n_j)}  \cdot\prod_{k=1}^\infty\frac{(1+q^{k-1})}{(1-q^{k})^N} =
\\
=q^{-\frac{(N-1)(2N-1)}{48N}}
\sum_{\alpha \in Q_{B_N}+\Lambda_N-\Lambda_0}
q^{\frac12(\alpha+\frac1{2N}\overline{\rho},\alpha+\frac1{2N}\overline{\rho})}
\prod_{k=1}^\infty\frac{(1+q^{k-1})}{(1-q^{k})^N}.
\ee	
where $\Lambda_N-\Lambda_0$ is the highest weight of the spinor representation of $SO(2N+1)$. The r.h.s. of these formulas are the characters of sums of nontwisted representations of $N$-component Heisenberg algebra with additional infinite-dimensional Clifford algebra (or real fermion). Another point of view that the r.h.s. are the characters of sums of representations of $W(B_N)$-algebra \cite{WBN}.

Finally, let us point out that for the root system $B_{2}^{(1)}=C_{2}^{(1)}$ (affine $B_2$) the dual roots system is $C_2^{(1),\vee}=D_3^{(2)}$ (affine twisted $D_3$).
\begin{equation}
\begin{aligned}
&\text{Simple roots: } \alpha_0=\delta-2e_1,\;\; \alpha_1=e_1-e_2,\;\; \alpha_2=2e_2.\\
&\text{Simple coroots: }\alpha^\vee_0=K-e_1^\vee,\;\; \alpha^\vee_1=e_1^\vee-e_2^\vee,\;\; \alpha^\vee_2=e_2^\vee.\\
&\text{Real coroots: } mK {\pm} e_1^\vee,\;\; mK {\pm} e_2^\vee,\;\; 2mK {\pm} e_1^\vee{\pm} e_2^\vee,\;\; 2mK {\pm} e_1^\vee{\mp} e_2^\vee,\, m\in \mathbb{Z}.\\
&\text{Imaginary coroots: } (2m-1)K\, \text{of multiplicity $1$},\, m\in \mathbb{Z}\, \\
&\qquad \qquad \qquad \qquad 2mK\, \text{of multiplicity $2$},\, m\in \mathbb{Z}\setminus\{0\}.\\
&\text{$\texttt{k}=1$ weights: } \Lambda_0,\;\; \Lambda_1=\epsilon_1+\Lambda_0,\;\; \Lambda_2=\Lambda_0+\epsilon_1+\epsilon_2 \\
&h=4,\;\rho=2e_1+e_2+3\Lambda_0,\;\overline{\rho}=\frac32e_1+\frac12e_2.
\end{aligned}
\end{equation}
The computation leads to result, coinciding with formulas \eqref{eq:BNden}, \eqref{eq:BNnum1}, \eqref{eq:BNnum2} for $N=2$. Though the root system here has a bit different combinatorial structure,
the fermionic construction is the same, using 5 real fermions.

\section{Exact conformal blocks of $W(\mf{so}(2N))$ twist fields}
\label{section:confblock}
\subsection{Operator algebra structure}

Now we are going to compute certain conformal blocks. We denote by $|\ddot{g}\rangle$ the highest weight vector of the representation of twisted Heisenberg $\widehat{\mathfrak{h}}$. We denote corresponding field by $\mc O_{\ddot{g}}$.

The fields $\mc O_{\ddot{g}}$ are primary fields for the W-algebra, so we compute conformal blocks for this algebra.\footnote{This is difficult to do using just W-algebra symmetry, the space of conformal blocks is infinite dimensional, they depend not only on the highest weights of external fields (so contrary to characters, the answer will depend on $\ddot{g}$, not just on its orbit under the action of $\rm N_G(\mathfrak{h})$).} But our theory has more symmetry, it contains fermions and bosons with nontrivial boundary conditions \eqref{eq:gentwist} and \eqref{eq:gentwistJ}. The presence of such operators provides additional restriction of the fusion of two fields
\begin{equation}
[\mc O_{\ddot{g}_1}][\mc O_{\ddot{g}_2}]=\sum [\mc O_{\ddot{h}}]
\end{equation}
First, the monodromy of the fused field should equal to the product of the monodromies $h=g_1g_2$. Second, we have a selection rule in terms of $r$-charges. Namely, for any zero mode in $\widehat{\mathfrak{h}}$, untwisted with respect to both $g_1, g_2$, corresponding $r$-charge for $\ddot{h}$ equals to the sum of $r$-charges of $g_1, g_2$. In particular, we have equality for total $r$-charges $\dot{h}=\dot{g}_1\dot{g}_2$. As an opposite example, if $g_1, g_2$ are both diagonal (this corresponds to the trivial element of the Weyl group), then all $r$-charges of $h$ equal to the sum of $r$-charges for $g_1$ and $g_2$.

In principle, such conformal block for twist fields can be studied for any $g \in G$, see \cite{GMfer} about their relation to the isomonodromic deformations. But here we restrict ourselves to the case $g \in \mr N_{G}(\mf h)$. If $g$ corresponds to nontrivial element of the Weyl group, then corresponding fields are special, for example in the case $G=GL(2)$ all fields, corresponding to transposition, have conformal dimension $\frac1{16}$. The corresponding conformal blocks were found by Al. Zamolodchikov in \cite{ZamAT}. Here we generalize his construction and give the answer in terms of the geometry of the branched cover.

\subsection{Global construction}

It has been shown in \cite{GMtw} that conformal block of the generic $W(\mf{gl}(N))$ twist fields is given by explicit formula, analogous to the famous Zamolodchikov's conformal blocks of the Virasoro twist fields with dimensions $\Delta = \frac1{16}$ \cite{ZamAT}.
To generalize the construction of \cite{GMtw} to all twist fields $\{\mathcal{O}_{\ddot g}|g\in \mr N_{G}(\mf h)\}$ considered in this paper, one needs to glue local data in the vicinity of all twist field to some global structure. We consider below such construction for $G=O(2N)$, since it can be entirely performed in terms of twisted bosons.

First, let us remind the local data in the vicinity of $\mc O_{\ddot g}(0)$ discussed already in Sect.~\ref{ss:twists}:
\begin{itemize}
\item $2l$-fold cover $z=\xi^{2l}$  with holomorphic involution $\sigma: \xi\mapsto-\xi$ without stable points except the twist field positions.
\item Fermionic field $\eta(\xi)$ with exotic OPE $\eta(\xi)\eta(\sigma(\xi'))\sim\frac1{\xi-\xi'}$. On the sheets, connected to each other by $[l,e^{2\pi i r}]_+$, one can identify $\eta(\xi)$ with the ordinary complex fermion $\psi(\xi)=\eta(\xi)$,
$\eta(\sigma(\xi))=\psi^*(\xi)$, in this case $\sigma$ permutes $\psi\leftrightarrow\psi^*$.
\item Bosonic field $J(z)=\left(\eta(\sigma(z))\eta(z)\right)$, which is antisymmetric $J(\sigma(z))=-J(z)$ under the action of involution $\sigma$, and has
first-order poles coming from zero-mode charges in the branch-points corresponding to cycles of type $[l,e^{2\pi i r}]_+$.
\end{itemize}
Now we want to compute spherical $2M$-point conformal block
\eq{
\mc G_0(q_1,\ldots, q_{2M})=\langle\mc O_{\ddot  g_1}(q_1)\mc O_{\ddot g_2}(q_2)\ldots \mc O_{\ddot  g_{2M-1}}(q_{2M-1})
\mc O_{\ddot g_{2M}}(q_{2M})\rangle_{\ddot h_1,\ldots, \ddot h_k},
\label{cnfbl}
}
where we fix intermediate fields $\mc O_{\ddot h_k}$ such that $h_k\in G$ are diagonal,  $g_{2k-1}g_{2k}=h_k$, the $r$-charges for $h_k$ should be compatible with the fusion $\mc O_{\ddot  g_{2k-1}}\mc O_{\ddot g_{2k}}(q_2)\sim\mc O_{\ddot h_k}$, and fusion of all $\mc O_{\ddot h_k}$ together equals to identity~\footnote{In principle, we may choose any monodromies, though in this way we will get complicated twisted
	representations in the intermediate channels, but as in \cite{GMtw} we restrict ourselves to simpler but still quite general case of pairwise inverse
	(up to diagonal factors $h_i$) monodromies.}.
In the discussion below we forget about fermion and consider only the twisted boson with current $J(z)$.
Now let us list the field-theoretic properties which fix this conformal block uniquely.

Let $g$ corresponds to the cycle $[l,\lambda]_+$,  denote by $|0\rangle_{\ddot g}$ the highest vector of the module of the twist-field $\mathcal{O}_{\ddot g}$. Then we have $J_{k/l>0}|0\rangle_{\ddot g}=0$. Therefore the most singular term of the 1-form $J(z)dz$ in the vicinity of the twist field $\mathcal{O}_{\ddot g}$ is the  simple pole
\eq{
J(z)dz\ \stackreb{z\to 0}{\sim}\ r\frac{dz}{z} + \ldots,
}
where the residue $r$ is the $r$-charge. Similarly, if $g$ corresponds to the cycle $[l]_-$, then $J(z)dz$ does not have pole in the vicinity of $\mc O_{\ddot g}$.

For two twist fields $\mc O_{\ddot g_{2k-1}}(z)$,$\mc O_{\ddot g_{2k}}(z')$  as above (i.e. corresponding to mutually-inverse elements of the corresponding Weyl group) the operator product expansion in the channel corresponding to $\ddot{h}_k$ has the form
\eq{
\mc O_{\ddot g_{2k-1}}(z)\mc O_{\ddot g_{2k}}(z')\stackreb{z\to z'}{\sim} (z-z')^{\Delta_{\ddot h_k}-\Delta_{\ddot g_{2k-1}}-\Delta_{\ddot g_{2k}}}V_{\ddot h}(z')+descendants}
where $V_{\ddot h_k}(z')=\mc O_{\ddot h_k}(z')$ is a field with fixed charges $\vec{a}\in \mathfrak{h}$~\footnote{Here $h_j=e^{2\pi ia_j}$, and since $h$ is diagonal, we may say that there is one to one correspondence between $\vec a$ and $\ddot h$: in $\ddot h$ we just fix the logarithm of $h$.} (we used another letter $V$ in order to stress that this is just exponent of the bosonic field). Hence
\eq{
\frac1{2\pi i}\Oint_{\mc C_{z,z'}^j}J(\xi)d\xi \mc O_{ \ddot g_{2k-1}}(z)\mc O_{\ddot g_{2k}}(z')=a_j\mc O_{
\ddot g_{2k-1}}(z)\mc O_{\ddot g_{2k}}(z')
}
where contour $\mc{C}_{z,z'}^j$ is the $j$-th preimage of the contour encircling two points $z,z'$ on the base. We identify such contours with the $\sf A$-cycles on the cover, and
corresponding $a$'s with the {\sf A}-periods of 1-form $J(z)dz$.

The standard OPE of two currents
\eq{
J(z)J(z')dz dz'\stackreb{z\to z'}{=}\frac{dz dz'}{(z-z')^2}+4\check{T}(z')+\ldots
\label{OPEJJ}
}
gives the stress-energy tensor
\be
T(z) = \Sum_{\pi_{2N}(\xi)=z}\check{T}(\xi)
\\
T(z)\mc O_{\ddot g}(0)=\frac{\Delta_{\ddot g}}{z^2}\mc O_{\ddot g}(0)+\frac{1}{z}\d\mc O_{\ddot g}(0)+\ldots
\label{TTc}
\ee
Non-standard coefficient (4 instead of 2) arises due to the involution $\sigma$.
Summarizing these facts we get:
\begin{itemize}
\item $2N$-sheet branched cover $\pi_{2N}:\Sigma\to\mathbb{P}^1$ with the branch points $\{q_1,\ldots,q_{2M}\}$
and ramification structure defined by the elements of the Weyl group, corresponding to $\{g_1, g_2, \ldots, g_{2M-1}, g_{2M}\}$. In particular,
$\Sigma$ is a disjoint union of two curves when all $\{g_i\}$ do not contain $[l]_-$ cycles.
\item Involution of this cover $\sigma:\Sigma\to\Sigma$ with the stable points coinciding with $[l_i]_-$ cycles
\begin{equation}
\begin{tikzcd}
\Sigma \arrow[r,"\pi_2"] \arrow[rr, bend left,"\pi_{2N}"] \arrow[loop left,"\sigma"]&\tilde\Sigma\arrow[r,"\pi_N"]&\mathbb{C}\mathbb{P}^1
\end{tikzcd}
\end{equation}
Projections and involution are shown on the commutative diagram: $\pi_{2N}=\pi_N\circ\pi_2$, $\pi_2\circ\sigma=\pi_2$.

\item Odd meromorphic differential $dS(\sigma(\xi))=-dS(\xi)$ with the poles in preimages of $q_i$ and residues given by corresponding $r$-charges.
\item Symmetric bidifferential $d\Omega(\xi,\xi')$, satisfying  $d\Omega(\sigma(\xi),\xi')=-d\Omega(\xi,\xi')$, with two poles:
    \eq{
d\Omega_2(\xi,\xi')\stackreb{\xi\to\xi'}{\sim}\frac{d\xi d\xi'}{(\xi-\xi')^2},\ \ \ \
d\Omega_2(\xi,\xi')\stackreb{\xi\to\sigma(\xi')}{\sim}-\frac{d\xi d\xi'}{(\xi-\sigma(\xi'))^2}
}
and vanishing $\sf A$-periods.
\end{itemize}
Using this data one can write the following formulas for two auxiliary correlators:
\eq{
\mc G_1(\xi|q_1,\ldots,q_{2M})=d\xi\langle J(\xi)\mc O_{\ddot g_1}(q_1)\mc O_{\ddot g_2}(q_2)\ldots \mc O_{\ddot  g_{2M-1}}(q_{2M-1})
\mc O_{\ddot g_{2M}}(q_{2M})\rangle\\
\mc G_2(\xi,\xi')=d\xi d\xi'\langle J(\xi)J(\xi')\mc O_{\ddot g_1}(q_1)\mc O_{\ddot g_2}(q_2)\ldots
\mc O_{ \ddot g_{2M-1}}(q_{2M-1}) \mc O_{\ddot g_{2M}}(q_{2M})\rangle
}
Their explicit expressions
\eq{
\mc G_1(\xi)\mc G_0^{-1}=dS(\xi),\ \ \ \
\mc G_2(\xi,\xi')\mc G_0^{-1}=dS(\xi)dS(\xi')+d\Omega_2(\xi,\xi')
\label{G12}
}
are fixed uniquely by their analytic behaviour. Now let us study in detail the structure of the curve $\Sigma$ in order to construct all these objects.

\subsection{ Curve with holomorphic involution}

Involution $\sigma$ defines the two-fold cover $\pi_2:\Sigma\to\tilde\Sigma$ with the total number
of branch points being $2K'=2\sum_{i=1}^MK_i'$, or exactly the total number of $[l]_-$ cycles in all elements $\{g_k\}$. Then the Riemann-Hurwitz formula $\chi(\Sigma) = 2\cdot\chi(\tilde{\Sigma}) - \# BP$ gives the genus
\eq{
\mathrm{g}(\Sigma)=2\mathrm{g}(\tilde\Sigma)+K'-1
}
A natural way to specify the ${\sf A}$-cycles on $\Sigma$ is the following \cite{Fay}: first to take ${\sf A}_1^{(1)},\ldots,{\sf A}_{\tilde{\mathrm{g}}}^{(1)}$,
${\sf A}_1^{(2)},\ldots,{\sf A}_{\tilde{\mathrm{g}}}^{(2)}$ on each copy of $\tilde{\Sigma}$, where $\tilde{\mathrm{g}} = \mathrm{g}(\tilde\Sigma)$; and second, all other
$\sf A$-cycles that correspond to the branch cuts of the cover, connecting the branch points of $\pi_2$: ${\sf A}_1^{(0)},\ldots,{\sf A}_{K'-1}^{(0)}$.
The action of the involution on these cycles is obviously given by
\eq{
\sigma({\sf A}_i^{(1)})={\sf A}_i^{(2)},\quad\sigma({\sf A}_i^{(2)})={\sf A}_i^{(1)},\ \ \ i=1,\ldots,\tilde{\rm g}
\\
\sigma({\sf A}_j^{(0)})=-{\sf A}_j^{(0)},\ \ \ j=1,\ldots,K'-1
\label{Ainvol}}
thus we have the decomposition of the real-valued first homology group into the even and odd parts
\eqs{
H_1(\Sigma,\mathbb R) & =H_1(\Sigma,\mathbb R)^{+}\oplus H_1(\Sigma,\mathbb R)^{-}\\
\dim H_1(\Sigma,\mathbb R) & ^{+}=\mathrm{g}(\tilde\Sigma) = \tilde{\mathrm{g}}\\
\dim H_1(\Sigma,\mathbb R) & ^-=\tilde{\mathrm{g}}+K'-1=\mathrm{g}_-
}
Compute now $\tilde{\mathrm{g}}=\mathrm{g}(\tilde\Sigma)$ using the Riemann-Hurwitz formula for the cover of $\mathbb{P}^1$. Let  $K=\sum_{i=1}^MK_i$ be the total number of $[l,e^{2\pi ir}]_+$-type cycles in all elements $\{g_{2k-1}\}$, as well as $K'$ serves for the type $[l']_-$. Then $\chi(\tilde{\Sigma}) = N\cdot\chi(\mathbb{P}^1) - \# BP$ gives (cf. with the formula (2.17) of \cite{GMtw})
\eq{
\tilde{\mathrm{g}}=1-N+\Sum_{i=1}^K(l_i-1)+\Sum_{i=1}^{K'}(l'_i-1)
}
so that
\eq{
\mathrm{g}_- = \tilde{\mathrm{g}}+K'-1 = \Sum_{i=1}^K(l_i-1)+\Sum_{i=1}^{K'}l'_i - N
\label{g-}
}
and
\be
\mathrm{g} =1-2N+2\Sum_{i=1}^K(l_i-1)+2\Sum_{i=1}^{K'}(l'_i-\half)
\ee
For our purposes the most essential is the odd part $H_1(\Sigma,\mathbb R)^-$ of the homology. One can see
these $\mathrm{g}_-$ $\sf A$-cycles explicitly as follows: two mutually inverse permutations of type $[l]_+$ produce $l$ pairs of $\sf A$-cycles
${\sf A}_i^{(1,2)}$ with constraints $\sum_i {\sf A}_i^{(1,2)}=0$. These cycles are permuted by $\sigma$ \rf{Ainvol}, so they actually form $l-1$ independent odd combinations, giving contribution to the r.h.s. of \rf{g-}. For two mutually inverse elements of the type $[l']_-$ one gets instead $2l'$ $\sf A$-cycles with constraint $\sum_i \sf A_i=0$, and with the action of involution $\sigma: \sf A_i\mapsto \sf A_{i+l'}$, giving $l'$ independent odd combinations $\{\sf A_i-
\sf A_{i+l'}\}$, arising in the r.h.s. of \rf{g-}, while the extra term $-N$ corresponds to the charge conservation in the infinity.

Hence, we got $\mathrm{g}_-$ odd $\sf A$-cycles, whose projections to $\mathbb{P}^1$ encircle pairs of the colliding twist fields
$\mc O_{\ddot g_{2k-1}}(q_{2k-1})\mc O_{\ddot g_{2k}}(q_{2k})$ for $k=1,\ldots,M$,  so that the integrals of
\eq{
\frac1{2\pi i}\oint_{{\sf A}_I} dS=a_I,\ \ \ I=1,\ldots,\mathrm{g}_-
}
give the W-charges in the intermediate channels of conformal block \rf{cnfbl}. Therefore $dS$ can be expanded
\eq{
dS=\Sum_{I=1}^{\mathrm{g}_-}a_I d\omega_I+\Sum_{i=1}^{2M}dS_{\bs r_i}
}
over the \emph{odd} holomorphic differentials and meromorphic differentials of the 3-rd kind
corresponding to the nonvanishing $r$-charges.

Now, for the bidifferential $d\Omega_2(\xi,\xi')$ one can write
\eq{
d\Omega_2(\xi,\xi')=K(\xi,\xi')-K(\sigma(\xi),\xi') = 2K(\xi,\xi')-\tilde K(\xi,\xi')
\label{OK}
}
where $K(\xi,\xi')$ is the canonical meromorphic bidifferential on $\Sigma$ (the double logarithmic derivative of the prime form, see \cite{Fay}), normalized on vanishing $\sf A$-periods in each of two variables, and
\eq{
\tilde K(\xi,\xi')=K(\xi,\xi')+K(\sigma(\xi),\xi')
\label{KSigt}
}
is actually a pullback of the canonical meromorphic bidifferential on $\tilde\Sigma$. Indeed, consider
\eq{
\delta K(\xi,\xi')=K(\xi,\xi')-K(\sigma(\xi),\sigma(\xi'))
}
which is already holomorphic at $\xi=\xi'$, and $\oint_{A_i} \delta K(\xi,\xi')=0$, since due to \rf{Ainvol} normalization conditions do not change under involution. Thus $\delta K(\xi,\xi')=0$ and the canonical bidifferential is $\sigma$-invariant
\eq{
K(\sigma(\xi),\sigma(\xi'))=K(\xi,\xi')
}
Moreover, since
\eq{
\tilde K(\xi,\xi') = \tilde K(\sigma(\xi),\xi')= \tilde K(\xi,\sigma(\xi'))
}
expression \rf{KSigt} actually defines the canonical bidifferential on $\tilde\Sigma$.

\subsection{Computation of the conformal block}

Now we use the technique from \cite{ZamAT,GMtw} to compute the conformal block \rf{cnfbl}. For the vacuum expectation value of the stress-energy tensor \rf{TTc}
one gets from \rf{G12}, \rf{OK}
\eq{
\langle T(z) \mc O_{\ddot g_1}(q_1)\mc O_{\ddot g_2}(q_2)\ldots \mc O_{\ddot g_{2M-1}}(q_{2M-1})
\mc O_{\ddot g_{2M}}(q_{2M})\rangle\mc G_0^{-1}=
\\
=\Sum_{\pi_{2N}(\xi)=z}t_z(\xi)-
\Sum_{\pi_N(\zeta)=z}\tilde t_z(\zeta)+\frac14\left(\frac{dS}{dz}\right)^2
\label{TtdS}
}
where $t_z$ and $\tilde t_z$ are the regularized parts of the bidifferentials $K$ and $\tilde K$ on diagonal in coordinate $z$:
\eqs{
t_z(\xi)d\xi^2&=\frac12\left(\lim_{\xi\to\xi'}K(\xi',\xi)-\frac{d\pi_{2N}(\xi) d\pi_{2N}(\xi')}{(\pi_{2N}(\xi')-\pi_{2N}(\xi))^2}\right)
\\
\tilde{t}_z(\zeta)d\zeta^2&=\frac12\left(\lim_{\zeta\to\zeta'}\tilde{K}(\zeta',\zeta)-\frac{d\pi_{N}(\zeta) d\pi_{N}(\zeta')}{(\pi_{N}(\zeta')-\pi_{N}(\zeta))^2}\right)
}
Expanding \rf{TtdS} at $z\to q_i$ one gets
\eqs{
\tilde{t}_z(\zeta)& \stackreb{z\to q_k}{=} \frac1{12}\{\zeta;z\} + reg. = \frac1{(z-q_k)^2}\frac{l^2-1}{24l^2} + reg.
\\
t_z(\xi)& \stackreb{z\to q_k}{=} \frac1{12}\{\xi;z\} + reg. = \frac1{(z-q_k)^2}\frac{4l'^2-1}{96l'^2} + reg.
}
in local co-ordinates $\xi^{2l'}=\zeta^l=z-q_k$, which gives for the conformal dimensions~\footnote{
The counting here works as
\be
t_z-\tilde{t}_z\rightarrow 2\sum_{j=1}^Kl_j\frac{l_j^2-1}{24l_j^2} - \sum_{j=1}^Kl_j\frac{l_j^2-1}{24l_j^2}
= \sum_{j=1}^K\frac{l_j^2-1}{24l_j}
\ee
for the $[l]_+$-cycles, and
\be
t_z-\tilde{t}_z\rightarrow \sum_{j=1}^{K'}2l'_j\frac{4l'^2_j-1}{96l'^2_j} - \sum_{j=1}^{K'}l_j'\frac{l_j'^2-1}{24l_j'^2}
= \sum_{j=1}^{K'}\frac{2l'^2_{j}+1}{48 l'_{j}}
\ee
for the $[l']_-$-cycles.
}
of the fields ${\mc O}_{\ddot g}$ (with generic $\mathfrak{o}(2N)$ twist field of the type \rf{geno2N})
\be
\Delta_{\ddot g}  = \sum_{j=1}^K \frac{l_j^2-1}{24l_j} +
\sum_{j=1}^{K'} \frac{2l'^2_{j}+1}{48 l'_{j}}
+\Sum_{i=1}^K\frac12 l_ir_i^2=\Delta_g^0+\Sum_{i=1}^K\frac12 l_ir_i^2\,,
\label{dimD}
\ee
where the last term in the r.h.s. comes from the expansion $dS \approx r_i\frac{dz}{z-q_i}+\ldots$. Without
contributions of the $r$-charges this formula is equivalent to \rf{dimD0}, \rf{dimD}.

From the first order poles we obtain
\eq{
\d_{q_k}\log\mc G_0(q_1,\ldots,q_{2M})=\Sum_{\pi_{2N}(\xi)=q_k}\Res t_z(\xi)d\xi - \Sum_{\pi_{N}(\zeta)=q_k}\Res \tilde t_z(\zeta)d\zeta +
\\
+\frac14\Sum_{\pi_{2N}(\xi)=q_k}\Res \frac{(dS)^2}{dz},\ \ \ \ k=1,\ldots,2M
}
This system of equations for conformal block is obviously solved, so that we can formulate:
\begin{theorem}
Conformal blocks \rf{cnfbl} for generic $W(\mathfrak{o}(2N))$ twist fields are given by
\eq{
\mc G_0(\bs a,\bs r, \bs q)=\tau_B(\Sigma|\bs q)\tau_B^{-1}(\tilde\Sigma|\bs q)\tau_{SW}(\bs a,\bs r,\bs q)
\label{th3}
}
where
\eq{
\d_{q_k}\log\tau_B(\Sigma|\bs q)=\Sum_{\pi_{2N}(\xi)=q_k}\Res t_z(\xi)d\xi\\
\d_{q_k}\log\tau_B(\tilde\Sigma|\bs q)=\Sum_{\pi_{N}(\zeta)=q_k}\Res \tilde t_z(\zeta)d\zeta
\\
k=1,\ldots,2M
\label{BK}
}
and
\be
\d_{q_k}\log\tau_{SW}(\bs a,\bs r,\bs q)=\frac14\Sum_{\pi_{2N}(\xi)=q_k}\Res \frac{(dS)^2}{dz},\ \ \ \ k=1,\ldots,2M
\\
\frac{\pd}{\pd{a_I}}\log\tau_{SW}=\oint_{{\sf B}_I}dS,\ \ \ \ {\sf A}_I\circ {\sf B}_J=\delta_{IJ},\ \ \ I,J=1,\ldots,\mathrm{g}_-
\label{SW}
\ee
\end{theorem}
Equations \rf{BK} define so-called Bergmann tau-functions \cite{KotKor1} for the curves $\Sigma$ and $\tilde{\Sigma}$ respectively, while the so-called Seiberg-Witten tau-function \rf{SW} can
be read literally from \cite{GMtw}
\be
\label{tQ}
\log\tau_{SW}(\bs a,\bs r,\bs q)=\frac14\Sum_{I,J=1}^{\mathrm{g}_-} a_I\mc T_{IJ} a_J+\frac12\Sum_{I=1}^{\mathrm{g}_-} a_IU_I(\bs r) + \frac14 Q(\bs r)
\ee
where $\mc T_{IJ}$ is the $\mathrm{g}_-\times \mathrm{g}_-$ ``odd block'' of the period matrix of $\Sigma$, or
the period matrix of corresponding Prym variety \cite{Fay}, the ``odd'' vector
\eq{
U_J(\bs r)=\Oint_{{\sf B}_J}d\Omega_{\bs r}=\Sum_{k,\alpha}r_k^{\alpha}{\mathcal A}_J(q_k^{\alpha}),\ \ \ J=1,\ldots,\mathrm{g}_-
\label{U}
}
where $q_k^\alpha$ are preimages of $q_k$, $r_k^\alpha$ --- corresponding $r$-charges, and
$\mathcal{A}_J(P)=\int^Pd\omega_J$ is the Abel map to the Jacobian of $\Sigma$. The last term in the r.h.s. of \rf{tQ} is given by
\eq{
Q(\bs r)=\sum_{q_i^\alpha\neq q_j^\beta}r_i^\alpha r_j^\beta\log\theta_*(\mathcal A(q_i^\alpha)-\mathcal A(q_j^\beta))-
\Sum_{q_i^\alpha}l_i^\alpha(r_i^\alpha)^2
\left.\log\frac{d(z(q)-q_i)^{1/{l_i^\alpha}}}{h_*^2(q)}\right|_{q=q_i^\alpha}
\label{Q}
}
where $\theta_*$ is some odd Riemann theta-function for the curve $\Sigma$, and
\eq{
h_*^2(z)=\Sum_{I=1}^{\rm g}\frac{\d\theta_*(0)}{\d Z_I}d\omega_I(z)
}

\paragraph{Remark:} In the general $N>2$ case conformal block constructed above is not fixed by conjugacy classes of twists and by charges in
the intermediate channels: it depends also on the geometry of the cover. This is a reminiscent of infinite-dimensional space of 3-point conformal blocks
in the general case, but unlike that case now we deal only with finite-dimensional space, which can be easily studied.

\subsection{Relation between $W(\mf{so}(2N))$ and $W(\mf{gl}(N))$ blocks}

It is interesting to compare the formulas from previous section with the formulas from \cite{GMtw} for the exact $W(\mf{gl}(N))$ conformal blocks. Since, as we already discussed
$W(\mf{so}(2N))\subset W(\mf{gl}(N))$, any vertex operator of the $W(\mf{gl}(N))$ algebra is a vertex operator of its subalgebra $W(\mf{so}(2N))$, and
it is clear from our construction that twist fields $\mathcal{O}_{\dot g}$ for the elements
$g\sim \Prod [l,e^{2\pi i r}]_+$, are also the twist fields for $W(\mf{gl}(N))$. Moreover,
the corresponding Verma modules, generated by $W(\mf{so}(2N))$ and by $W(\mf{gl}(N))$, actually coincide~\footnote{These two modules
coincide due to dimensional argument: they are both irreducible and have the same characters. Irreducibility follows from the fact that the
null-vector condition can be written as $\left(\alpha,\frac{\log g}{2\pi i}\right)\in\mathbb Z$ for a simple root $\alpha$, and for generic $r$'s it is violated, see also comments in Sect.~\ref{ssec:W:decom}.},
and it means that corresponding conformal blocks of such fields in these two theories should coincide
as well.

Indeed, in such a case $\Sigma=\tilde\Sigma\sqcup\tilde\Sigma$, and therefore $K(\xi,\xi')=0$ if $\xi'$, $\xi$ are on the different components, and $K(\xi,\xi')=\tilde
K(\xi,\xi')$ if they are on the same component, hence
\eq{
t_z(z)=2\tilde t_z(z)
}
For holomorphic and meromorphic differentials, one has in this case in the natural basis
\eq{
a_I = \Oint_{{\sf A}_I^{(1)}} dS=-\Oint_{{\sf A}_I^{(2)}} dS,\ \ \ \ \ \ I=1,\ldots,\tilde{\rm g}
\\
r_k^\alpha = \Res_{q_k^\alpha} dS=-\Res_{\sigma(q_k^\alpha)} dS
}
for the preimages $\{q_k^\alpha\}$ on $\tilde{\Sigma}$, and the period matrix of $\Sigma$ consists of two nonzero $\tilde{\rm g}\times \tilde{\rm g}$ blocks:
\eq{
\mc T^{(11)}=\mc T^{(22)}=\tilde{\mc T}
}
Under such conditions formula \rf{th3} turns into
\eq{
\mc G_0(\bs a,\bs r, \bs q)=\tau_B(\tilde\Sigma|\bs q)\tilde{\tau}_{SW}(\bs a,\bs r,\bs q)
}
where
\eq{
\log\tilde{\tau}_{SW}(\bs a,\bs r,\bs q)=\half\Sum_{I,J=1}^{\tilde{\mathrm{g}}} a_I\tilde{\mc T}_{IJ} a_J+\Sum_{I=1}^{\tilde{\mathrm{g}}} a_I\tilde U_I(\bs r) + \half \tilde Q(\bs r)
}
with corresponding obvious modifications of the formulas \rf{U} and \rf{Q}, which gives exactly the
$W(\mf{gl}(N))$ conformal block in terms of the data on smaller curve $\tilde{\Sigma}$.

\section{Conclusion}

In this paper we have considered the twist fields for W-algebras with integer Virasoro central charges, which are labeled by the conjugacy classes in the Cartan normalizers $\mr N_{G}(\mf h)$ of corresponding Lie groups and some extra discrete data. In addition to the most common $W_N$-algebras, corresponding to $A$-series (or $W(\mf{gl}(N))=W_N\oplus\sf H$, coming from $G=GL(N)$), we have extended this construction for the $G=O(n)$ case, which includes in addition to $D$-series the non simply-laced
$B$-case with the half-integer Virasoro central charge.

In terms of two-dimensional conformal field theory our construction is based on the free-field representation, where generalization to the $D$-series and $B$-series exploits the theory of real fermions, which in the odd $B$-case cannot be fully bosonized, so that in addition to the modules of twisted Heisenberg algebra one has to take into account those of the infinite-dimensional Clifford algebra. This construction produces
representations of the W-algebras (that are at the same time the twisted representations of corresponding KM algebras), which can be decomposed
further into the Verma modules. To find this decomposition we have computed the characters
of twisted representations, using two alternative methods.

The first one comes from bosonization of the W-algebra or the corresponding KM algebra at level one.
Dependently on particular element from $\mr N_{G}(\mf h)$ it identifies the representation space with a collection of the Fock modules for untwisted or twisted bosons. The essential new phenomenon, which appears
in the case of orthogonal groups, is presence of different $[l]_-$ cycles in $g\in\mr N_{G}(\mf h)$ and
necessity to use in such cases the ``exotic'' bosonization for the Ramond-type fermions with non-local OPE
on the cover.
Alternative method for computation of the characters uses pure algebraic construction of the twisted Kac-Moody algebras and the Weyl-Kac formula in principal specialization.

Since there are the elements $g_1, g_2$ not conjugated in $\mr N_{G}(\mf h)$, but conjugated in $G$, two different constructions with $g_1$ and $g_2$ give formally different formulations of the same representation. The computation of corresponding characters $\chi_{\dot g_1}(q)$ and $\chi_{\dot g_2}(q)$
leads therefore
to quite simple but nontrivial identities for the corresponding lattice theta-functions, $\chi_{\dot g_1}(q)=\chi_{\dot g_2}(q)$, which have been also proven
by direct methods.

We have also derived an exact formula for generic conformal block of the twist fields in $D$-case, which
directly generalizes the corresponding construction for common $W_N$-algebra. The result, as is usual for Zamolodchikov's exact conformal block, is expressed in terms of geometry of the covering curve (here with extra
involution), and can be factorized into the classical ``Seiberg-Witten'' part, totally determined by the period matrix of the corresponding Prym variety, and the quasiclassical correction, expressed now in terms
of two different canonical bi-differentials. In order to expand this method for the $B$-case one has to learn more about the theory of ``exotic fermions'' on Riemann surfaces, probably along the lines of \cite{DiFr,DVVc1}, and we postpone this for a separate publication.

Another set of open problems is obviously related with generalization to other series and twisted fields related with external automorphisms. Here only the $E$-cases seem to be straightforward, since standard bosonization can be immediately applied in the simply-laced case, and there should not be many problems with the fermionic construction. However, it is not easy to predict what happens in the situation when Kac-Moody algebras at level $\texttt{k}=1$ have fractional central charges, and the direct application of the methods developed in this paper is probably impossible. It is still not very clear, what is the role of these exact conformal blocks in the context of multi-dimensional supersymmetric gauge theories (not even thinking about other possible multidimensional applications of the W-algebras!), since generally there is no Nekrasov combinatorial representation in most of the cases.

Finally, there is an interesting question of possible generalization of our approach to the twisted representations with $\texttt{k}\neq 1$, which has been already considered in \cite{FSS}. Some overlap
of our formulas with Sect.~8 of this paper suggests that such generalization could exist. We hope to return to all these issues in the future.

\section*{Acknowledgements}
We are grateful to D.~Adler, B.~Feigin, I.~Frenkel, A.~Ilyina, I.~Krichever, O.~Lisovyy, E.~Opdam, and  N.~Sakharova for illuminating discussions, and to the organizers of ``5th workshop on Combinatorics of moduli spaces'' in Moscow, ``CQISS-16'' at the Chern Institute of Nankai University, ``Random geometry and physics'' in Paris and ``Progress in quantum field theory and string theory II'' in Osaka, where the preliminary results of this
work have been reported. The work of PG and AM was partially supported by the RSF grant No. 16-11-10160, in particular the main results of Sect.~\ref{sec:characters} and the construction
from Sect.~\ref{section:confblock} have been obtained using support of Russian Science Foundation,
the work of AM has been also partially supported by RFBR grant 17-01-00585 and
the RFBR/JSPS joint project 17-51-50051.
This work has been also funded by the  Russian Academic Excellence Project
'5-100'.

PG and AM also like to thank the KdV Institute of
the University of Amsterdam, where essential part of this work has been done, and especially G.~Helminck, for the warm hospitality.
MB and PG are Young Russian Mathematics award winners and would like to thank its sponsors and jury.

\newpage

\section*{Appendix}

\appendix

\section{Identities for lattice $\Theta$-functions
\label{app:thetaid}}

Here we present few rigorously proved identities, used to verify representation-theoretic considerations at the level of computations of the characters.

\subsection{First identity for $A_{N-1}$ and $D_N$ $\Theta$-functions}

One can describe the lattices $A_{N-1}$, $D_N$ and $D_N'$ in a similar way:
\eqs{
A_{N-1}=&\{k_1,\ldots,k_N\left|\Sum_{i=1}^N k_i=0\}\right.\\
D_{N}=&\{k_1,\ldots,k_N\left|\Sum_{i=1}^N k_i\in 2\mathbb Z\}\right.\\
D_{N}'=&\{k_1,\ldots,k_N\left|\Sum_{i=1}^N k_i\in 2\mathbb Z+1\}\right.
\label{ADD}
}
The last lattice is actually just $D_N$ lattice, but shifted by the vector $(1,0,\ldots,0)$. So all these definitions can be rewritten as
\eq{
L_S=\{k_1,\ldots,k_N\left|\Sum_{i=1}^N k_i\in S\}\right.
}
where $S\subseteq \mathbb Z$: in our cases it should be chosen to be $\{0\}$, $2\mathbb Z$, and $2\mathbb Z+1$, respectively. Notice also that for $S=\mathbb Z$ we get the simplest $B_N$ lattice.

By definition
\eq{
\Theta_{L_S}(\vec v;q)=\Sum_{k_1+\ldots k_N\in S}q^{\frac12(\vec v+\vec k)^2}
}
For our purposes we need this function computed for the vector
\eq{
\vec v=(r_1+\frac{l_1-1}{2l_1},r_1+\frac{l_1-3}{2l_1},\ldots,r_1+\frac{1-l_1}{2l_1})\oplus \\\oplus
(r_2+\frac{l_2-1}{2l_2},r_2+\frac{l_2-3}{2l_2},\ldots,r_2+\frac{1-l_2}{2l_2})\oplus\ldots\oplus
\\
\oplus(r_K+\frac{l_K-1}{2l_K},r_K+\frac{l_K-3}{2l_K},\ldots,r_K+\frac{1-l_K}{2l_K})
\label{vecv}}
where $l_1+\ldots+l_K=N$.
Let us parametrize vector $\vec k$ as follows:
\eq{
\vec k=(n_1,\ldots, n_1)\oplus\ldots\oplus(n_K,\ldots,n_K)+\omega^{(l_1)}_{a_1}\oplus\ldots\oplus\omega^{(l_K)}_{a_K}+\\
+(\frac{a_1}{l_1},\ldots, \frac{a_1}{l_1})\oplus\ldots
\oplus(\frac{a_K}{l_K},\ldots,\frac{a_K}{l_K})+\vec m_1\oplus\ldots
\oplus\vec m_K
\label{kpar}
}
where $\vec m_i\in A_{l_i-1}$, and
\eq{
\omega_{a}^{(l)}=(\frac{l-a}{l},\ldots,\frac{l-a}{l},-\frac{a}{l},\ldots,-\frac{a}{l})
}
so that the first number is repeated $a$ times, whereas the second one $l-a$ times.
Hence, vectors $\vec k\in L_S$ are parametrized by vectors $\{\vec m_i\in A_{l_i-1}\}$ and integer numbers
$\{ n_i\in \mathbb{Z}; a_i\in \mathbb{Z}/l_i\mathbb{Z}\}$, restricted by
\eq{
\Sum_{i=1}^K (n_il_i+a_i)\in S
}
The algorithm of decomposition \rf{kpar} works as follows: first we sum up all components of $\vec{k}$ inside each cycle --- each number divided by $l_i$ gives $n_i$ with a remainder $a_i$. Subtracting $(n_i,\ldots,n_i)+\omega_{a_i}^{(l_i)}$, we are left with the vectors $\{\vec m_i\}$ with vanishing sums of components.

Now it is easy to see that
\eq{
\Theta(\vec v+\omega^{(l_1)}_{a_1}\oplus\omega^{(l_2)}_{a_2}\oplus\ldots\oplus\omega^{(l_K)}_{a_K};q)=\Theta(\vec v;q)
}
which follows from the fact that $\Theta(\vec v;q)=\Theta(\sigma(\vec v);q)$, where $\sigma$ is a permutation. For example, take $\sigma_a$ to be $a$-th power of the cyclic permutation, then:
\eq{
\sigma_a\left(\frac{1-l}{2l},\ldots,\frac{l-1}{2l}\right)=
\left(\frac{l+1-2a}{2l},\frac{l+3-2a}{2l},\ldots,\frac{l-1}{2l},\frac{1-l}{2l},\ldots,\frac{l-1-2a}{2l}\right)=
\\
= \left(\frac{1-l}{2l},\ldots,\frac{l-1}{2l}\right)+\omega_a^{(l)}
}
and therefore any vector $\vec v+\omega^{(l_1)}_{a_1}\oplus\omega^{(l_1)}_{a_2}\oplus\ldots\oplus\omega^{(l_K)}_{a_K}$ can be obtained by several
permutations of components of  $\vec v$, so the corresponding $\Theta$-functions are equal. Thus
\eq{
\Theta_{L_S}(\vec v;q)=\Sum_{\substack{\Sum_{i=1}^K(n_il_i+a_i)\in S\\\vec m_i\in Q_{A_{l_i-1}}}}q^{\frac12(
\vec{v} + \vec m_1\oplus\ldots\oplus\vec m_K+
(n_1+\frac{a_1}{l_1},\ldots,n_1+\frac{a_1}{l_1})\oplus\ldots\oplus(n_K+\frac{a_K}{l_K},\ldots,n_K+\frac{a_K}{l_K}))^2}
}
turns into the sum over several orthogonal sublattices
\eq{
\Theta_{L_S}(\vec v;q)=\Sum_{\vec m_i\in Q_{A_{l_i-1}}}
q^{\frac12(\hat\rho^{(l_1)}\oplus\ldots\oplus\hat\rho^{(l_K)}+\vec m_1\oplus\ldots\oplus\vec m_K)^2}\cdot
\Sum_{\Sum_{i=1}^K(n_il_i+a_i)\in S}q^{\frac12\Sum_{i=1}^K l_i(n_i+\frac{a_i}{l_i}+r_i)^2}=\\
=\Prod_{i=1}^K\Theta_{A_{l_i-1}}(\hat\rho^{(l_i)};q)\cdot\Sum_{n'_1+\ldots+n'_K\in S}q^{\Sum_{i=1}^K\frac1{2l_i}(n_i'+r_il_i)^2}
\label{identity0}}
where
\eq{
\hat\rho^{(l)}=(\frac{l-1}{2l},\frac{l-3}{2l},\ldots,\frac{1-l}{2l})
\label{rhol}
}
One can identify the last factor in the r.h.s. with the contribution of zero modes, related to the $r$-charges \cite{GMtw}.

\subsection{A product formula for $A_{N-1}$ $\Theta$-functions}

Apply \rf{identity0} to the simplest case of $\Theta_{B_N}(\hat\rho^{(N)};q)$ with $S=\mathbb Z$
\eq{
\Theta_{B_N}(\hat\rho^{(N)};q)=\Theta_{A_{N-1}}(\hat\rho^{(N)};q)
\cdot\Sum_{n\in\mathbb Z}q^{\frac{n^2}{2N}}
\label{thetaAB}
}
Using the definition \rf{rhol} and the Jacobi triple product formula we get
\eq{
\Theta_{B_N}(\hat\rho^{(N)};q)=q^{\frac{N^2-1}{24N}}\prod_{a=0}^{N-1}\Sum_{k\in\mathbb Z}q^{\frac{k^2}2 + \frac{N-1-2a}{2N}k}=
\\
= q^{\frac{N^2-1}{24N}}\Prod_{k=1}^\infty(1+q^{\frac1N(k-\frac12)})^2\Prod_{n=1}^\infty(1-q^n)^N
}
as well as
\eq{
\Sum_{n\in\mathbb Z}q^{\frac{n^2}{2N}}=\Prod_{k=1}^\infty(1+q^{\frac1N(k-\frac12)})^2\Prod_{n=1}^\infty(1-q^{\frac nN})
}
Substituting this into \rf{thetaAB} one obtains
\eq{
\Theta_{A_{N-1}}(\hat\rho^{(N)};q)=q^{\frac{N^2-1}{24N}}\frac{\Prod_{k=1}^\infty(1-q^k)^N}{\Prod_{k=1}^\infty(1-q^{\frac kN})}=\frac{\eta(q)^N}{
\eta(q^{\frac1N})}
\label{ANprod}
}
or the product formula \cite{Macdonald:1972} for $\Theta_{A_{N-1}}(\hat\rho^{(N)};q)$, where the r.h.s. is expressed in terms of the Dedekind functions.
Substituting this into \rf{identity0} we get it in its final form
\eq{
\Theta_{L_S}(\vec v;q) =
\Sum_{k_1+\ldots+k_N\in S}q^{\frac12\Sum_{i=1}^N(v_i+k_i)^2}=
\Prod_{i=1}^K\frac{\eta(q)^{l_i}}{\eta(q^{\frac1{l_i}})}\cdot\Sum_{n_1+\ldots+n_K\in S}q^{\Sum_{i=1}^K\frac1{2l_i}(n_i+l_ir_i)^2}
\label{identity1}
}

\subsection{An identity for $D_N$ and $B_N$ $\Theta$-functions}

Here we show how $\Theta_{D_N}(\vec v_\ast;q)$ can be simplified if $\vec v_\ast$ contains at least one component $\frac12$. One has then
\eq{
\Theta_{D_N}(\vec v_\ast;q) = \Theta_{D_N}((\half,v_2,\ldots,v_n);q)=\Sum_{k_1+\ldots+k_n\in2\mathbb Z}q^{\frac12(\vec v_\ast+\vec k)^2}=\\=
\Theta_{D_N}((-\half,v_2,\ldots,v_n);q)=\Theta_{D_n}(\vec v_\ast-(1,0,\ldots,0);q)
\label{BD}
}
Since for the lattices $D_N\sqcup\{D_N-(1,0,\ldots,0)\}=B_N$, it follows from \rf{BD} that
\eq{
\Theta_{D_N}(\vec v_\ast;q)=\half\Theta_{B_N}(\vec v_\ast;q)
\label{identity2}
}

\section{Exotic bosonizations}

Here we present some details of the bosonization procedures used in the main text.

\subsection{$NS\times R$
\label{NSR}}

Consider first construction \cite{Bernard,BBT} relating pair (of $NS$ and $R$!) fermions to a twisted boson~\footnote{It is more convenient to use in
this section coordinate $\xi=\sqrt{t}$, so analytic continuation in $t$ around $0$ maps $\xi$ to $-\xi$. }
\eq{
\tilde\phi(t)=i\Sum_{r\in \mathbb Z +\frac12}\frac{J_r}{rt^r}=i\sqrt{2}\Sum_{n\in \mathbb Z}\frac{a_{2n+1}}{(2n+1)\xi^{2n+1}} = \phi(\xi)
}
with differently normalized oscillator modes $[a_M,a_N]=M\delta_{M+N,0}$ ($M,N\in 2\mathbb{Z}+1$). Compute the correlator
\eq{
-\langle\phi(\xi)\phi(\zeta)\rangle=2\sum \langle a_{2n+1}a_{2m+1}\rangle \xi^{-2n-1}\zeta^{-2m-1}=
-2\Sum_{n=0}^\infty\frac{(\zeta/\xi)^{2n+1}}{2n+1}=\\
=2\log\left(1-\frac \zeta\xi\right)-\log\left(1-\frac {\zeta^2}{\xi^2}\right)=-\log\frac{\xi+\zeta}{\xi-\zeta}=-[\phi_+(\xi),\phi_-(\zeta)]
}
assuming $|\xi|>|\zeta|$. Now introduce
\eq{
\hat\eta(\xi)=\frac1{\sqrt2}:e^{i\phi(\xi)}:=\frac1{\sqrt2}e^{i\phi_-(\xi)}e^{i\phi_+(\xi)}
}
so that for $|\xi|>|\zeta|$
\eq{
\hat\eta(\xi)\hat\eta(\zeta)=\frac12e^{-(\phi_-(\xi)+\phi_-(\zeta))}e^{i(\phi_+(\xi)+\phi_+(\zeta))}e^{-[\phi_+(\xi),\phi_-(\zeta)]}=\\=
\frac12:e^{i(\phi(\xi)+\phi(\zeta))}:\frac{\xi-\zeta}{\xi+\zeta}
}
while for $|\xi|<|\zeta|$
\eq{
\hat\eta(\zeta)\hat\eta(\xi)=\frac12:e^{i(\phi(\xi)+\phi(\zeta))}:\frac{\zeta-\xi}{\xi+\zeta}
}
It means that OPE of the $\hat{\eta}$-fields has fermionic nature:
\eq{
\hat\eta(\xi)\hat\eta(-\zeta)=\frac12\frac{\xi+\zeta}{\xi-\zeta}:e^{(\phi(\xi)-\phi(\zeta))}:\sim\frac12\frac{\xi+\zeta}{\xi-\zeta}+reg.\sim\frac{\zeta}{\xi-\zeta}+reg.
}
and in the anticommutator of components $\hat\eta(\xi)=\Sum_{k\in\mathbb Z}\frac{\eta_k}{\xi^k}$
\eq{
\{\eta_k,(-1)^l\eta_l\}=\oint \zeta^{l-1}d\zeta\oint_\zeta\frac{\zeta}{\xi-\zeta}\xi^{k-1}d\xi=\delta_{k+l,0}
}
one gets unusual sign factor.

It is interesting to point our that the Ramond zero mode $\eta_0^2=\half$ has bosonic representation
\eq{
\sqrt2 \eta_0=\oint\frac{d\xi}\xi e^{i\phi_-(\xi)}e^{i\phi_+(\xi)}=
\\
=1-2a_{-1}a_1+a_{-1}^2a_1^2-\frac29(a_{-3}+a_{-1}^3)(a_3+a_1^3)+\ldots
}
For example, the action of this operator on low-level vectors gives
\be
\sqrt2\eta_0\cdot|0\rangle=|0\rangle,\ \ \
\sqrt2\eta_0\cdot a_{-1}|0\rangle=-a_{-1}|0\rangle,\ \ \
\sqrt2\eta_0\cdot a_{-1}^2|0\rangle=a_{-1}^2|0\rangle\\
\sqrt2\eta_0\cdot a_{-3}|0\rangle=\frac13a_{-3}|0\rangle-\frac23a_{-1}^3|0\rangle,\ \ \
\sqrt2\eta_0\cdot a_{-1}^3|0\rangle=-\frac43a_{-3}|0\rangle-\frac13a_{-1}^3|0\rangle
\ee
Here in the second line one gets the matrix $\frac13\bpm1&-2\\-4&-1\epm$ with the eigenvalues $\pm1$.
We also have
\eq{
\eta_0\eta_k=-\eta_k\eta_0,\quad k\neq 0
}
so one can identify $\sqrt2\eta_0=(-1)^{F-F_0}$, where $F$ is the fermionic parity. Generally, algebra, generated by $\{\eta_k\}$,
has two representations with the vacua $|0\rangle_\pm$, such that $\eta_0|0\rangle_\pm=\pm|0\rangle_\pm$. One can also take direct sum of such representations:
bosonization formula in this representation looks as
\eq{
\hat\eta(\xi)=\frac{\sigma_1}{\sqrt 2}e^{i\phi_-(\xi)}e^{i\phi_+(\xi)}
}
Existence of this bosonization at the level of characters gives us obvious identity
\eq{
\prod_{k=0}^\infty\frac1{1-q^{2k+1}}=\prod_{k=1}^\infty(1+q^k)
\label{simpid}
}
Notice that above consideration actually concerns $R$ and $NS$ fermions because one can construct two combinations
\be
\frac1{\sqrt{2}}\left(\hat\eta(z)-\hat\eta(-z)\right) = \sum_{p\in \mathbb{Z}+\frac12} \frac{\eta_{2p}}{z^{p}} =i\hat\psi_{NS}(z)
\\
\frac1{\sqrt{2}}\left(\hat\eta(z)+\hat\eta(-z)\right) = \sum_{n\in \mathbb{Z}} \frac{\eta_{2n}}{z^{n}} = \hat\psi_{R}(z)
\ee
then
\be
J(z)=\frac1z\left(\hat\psi^*(\sqrt{z})\hat\psi(\sqrt{z})\right)=i\psi_{NS}(z)\psi_{R}(z) = \sum_{p\in \mathbb{Z}+\frac12}
\frac{J_{p}}{z^{p+1}}
\\
J_p = i\sum_{n+q=p}\eta_{2q}\eta_{2n}
\ee
here $t^{-\frac12}\hat\psi_{NS}(\sqrt{t})$ and $t^{-\frac12}\hat\psi_{R}(\sqrt{t})$ are usual Ramond and Neveu-Schwarz fermions.

Here we consider fermion corresponding to the branch point of type $[l]_-$. This means that we should have
\eq{
\eta(z)\eta(\sigma(w))\sim\frac1{z-w}\,,
}
and such monodromy that $\eta(e^{4\pi il} z)=\pm\eta(z)$. Let us use the construction form \rf{NSR}
\eq{
\eta(z)=\frac{z^{-\frac12}}{\sqrt{2l}}\hat\eta(z^{\frac1{2l}})
}
Therefore
\eq{
\eta(z)\eta(\sigma(w))\sim\frac{z^{-\frac12}w^{-\frac12}}{2l}\frac{w^{\frac1{2l}}}{z^{\frac1{2l}}-w^{\frac1{2l}}}\sim\frac1{z-w}
}
So final construction states that one should have
\eq{
\eta(z)=\sigma_1\frac{z^{-\frac12}}{2\sqrt{l}}e^{i\phi_-\left(z^{\frac1{2l}}\right)}e^{i\phi_+\left(z^{\frac1{2l}}\right)}
\label{bosonizationeta}
}

\subsection{$R\times R$}

Let us take two Ramond fermions $\psi^{(1)}$, $\psi^{(2)}$ and introduce
\eq{
\psi(z)=\frac1{\sqrt2}\left(\psi^{(1)}(z)+i\psi^{(2)}(z)\right)=\Sum_{n\in\mathbb Z}\frac{\psi_n}{z^{n+\frac12}}
\\
\psi^*(z)=\frac1{\sqrt2}\left(\psi^{(1)}(z)-i\psi^{(2)}(z)\right)=\Sum_{n\in\mathbb Z}\frac{\psi_n^*}{z^{n+\frac12}}
\label{RR}
}
Since there are two zero modes $\psi_0^*$ and $\psi_0$, one expects to have four vacua $|0\rangle$, $\psi_0|0\rangle$, $\psi_0^*|0\rangle$,
$\psi_0^*\psi_0|0\rangle$.

We can mimic expansion \rf{RR} using fractional powers
\eq{
\psi(z)=\Sum_{p\in\mathbb Z+\frac12}\frac{\psi_{NS,p}}{z^{p+\frac12+\sigma}},\ \ \ \
\psi^*(z)=\Sum_{p\in\mathbb Z+\frac12}\frac{\psi^*_{NS,p}}{z^{p+\frac12-\sigma}}
}
with $\sigma=\frac12$, i.e. $\psi_{n}=\psi_{NS,n-\frac12}$ and $\psi^*_{n}=\psi^*_{NS,n+\frac12}$. It means that after standard bosonization
\be
\psi(z)=e^{-i\phi_-(z)}e^{-i\phi_+(z)}e^{-Q}z^{-J_0},\ \ \ \
\psi^*(z)=e^{i\phi_-(z)}e^{i\phi_+(z)}e^{Q}z^{J_0}\\
J_0|0\rangle =\sigma|0\rangle =\half|0\rangle
\ee
one gets $\psi_0^*|0\rangle=0$, and only one half of the vacuum states survive. To identify this representation with something well-known, consider the eigenvectors $\sqrt2\psi_0^{(1)}|0\rangle_\pm = \pm |0\rangle_\pm $ of $\sqrt2\psi_0^{(1)}=\psi_0+\psi_0^*$:
\eq{
|0\rangle_+=\frac1{\sqrt2}(|0\rangle+\psi_0|0\rangle),\ \ \
|0\rangle_-=\frac i{\sqrt2}(|0\rangle-\psi_0|0\rangle)
}
Acting by $\sqrt2\psi_0^{(2)}=i(\psi_0^*-\psi_0)$ one gets
\be
\sqrt2\psi_0^{(2)}|0\rangle_+=|0\rangle_-,\ \ \ \
\sqrt2\psi_0^{(2)}|0\rangle_-=|0\rangle_+
\ee
The character of such module is given by
\eq{
2\prod_{k=1}^\infty(1+q^k)^2=q^{-\frac18}\frac{\Sum_{n\in\mathbb Z} q^{\frac12(n+\frac12)^2}}{\Prod_{k=1}^\infty(1-q^k)}
\label{JacobiProd1}
}
where in the l.h.s. we have two Ramond fermions with two vacuum states, whereas the r.h.s. corresponds to the sum over bosonic modules with half-integer vacuum $J_0$ charges. This formula is a simple consequence of the Jacobi triple product identity. Analogously we have similar formula for the bosonization
of $NS\times NS$ fermions
\eq{
\prod_{k=0}^\infty(1+q^{\frac12+k})^2=\frac{\Sum_{n\in\mathbb Z}q^{\frac12 n^2}}{\prod_{k=1}^\infty(1-q^k)}
\label{JacobiProd2}
}
It is the consequence of Jacobi triple product identity as well.
\subsection{$l$ twisted charged fermions}
\label{app:gln}

For the twisted boson
\eq{
i\phi(z)=-\Sum_{n\neq 0}\frac{J_{n/l}}{nz^{n/l}}+\frac1lJ_0\log z+Q
}
with
\eqs{
\left[J_{n/l},J_{m/l}\right]=n\delta_{n+m,0}\ \ \
[J_0,Q]=1
}
one has for $|z|>|w|$
\eq{
\left[\phi_+(z)-\frac ilJ_0\log z,\phi_-(w)-iQ\right]=\Sum_{n>0}\frac{z^{-n/l}w^{n/l}}{n}-\frac1l\log z=-\log\left(z^{1/l}-w^{1/l}\right)
}
where
\eqs{
i\phi_+(z)=-\Sum_{n>0}\frac{J_{n/l}}{nz^{n/l}}\ \ \ \
i\phi_-(z)=-\Sum_{n<0}\frac{J_{n/l}}{nz^{n/l}}
}
Define two operators
\eqs{
\hat\psi^*(z)=z^{\frac1{2l}}:e^{i\phi(z)}&:
=z^{\frac1{2l}}e^{i\phi_-(z)}e^{i\phi_+(z)}e^Qz^{J_0/l}
\\
\hat\psi(z)=z^{\frac1{2l}}:e^{-i\phi(z)}&:
=z^{\frac1{2l}}e^{-i\phi_-(z)}e^{-i\phi_+(z)}e^{-Q}z^{-J_0/l}
\label{definition1}
}
with the OPE
\eq{
\hat\psi^*(z)\hat\psi(w) =
\frac{(zw)^{\frac1{2l}}}{z^{1/l}-w^{1/l}}:e^{i\phi(z)-i\phi(w)}: =
\\
=
\frac{(zw)^{\frac1{2l}}}{z^{1/l}-w^{1/l}}e^{i\phi_+(z)-i\phi_+(w)}e^{i\phi_-(z)-i\phi_-(w)}
\left(\frac{z}{w}\right)^{J_0/l}
\label{opehat}
}
Then for the modes of their expansion
\be
\hat\psi^*(z)=\Sum_{k\in\frac12+\mathbb Z}\frac{\psi^*_{k/l}}{z^{k/l}},\hspace{1cm}
\hat\psi(z)=\Sum_{k\in\frac12+\mathbb Z}\frac{\psi_{k/l}}{z^{k/l}}
\label{modes}
\ee
one gets canonical anticommutation relations
\eq{
\{\psi_a^*,\psi_b\}=\delta_{a+b,0}
}
Now one can express the $l$-component fermions in terms of a single twisted boson
\eq{
\psi^*_\alpha(z)=\frac1{\sqrt l}z^{-\frac12}\hat\psi^*(e^{2\pi i\alpha} z),\ \ \ \
\psi_\alpha(z)=\frac1{\sqrt l}z^{-\frac12}\hat\psi(e^{2\pi i\alpha} z),\ \ \ \ \alpha\in \mathbb{Z}/l\mathbb{Z}
\label{definition2}
}
and it follows from \rf{opehat}, that their OPE is indeed
\eq{
\psi^*_\alpha(z)\psi_\beta(w)\ \stackreb{z\to w}{=}\ \frac{\delta_{\alpha\beta}}{z-w}+reg.
}
The stress-energy tensor and $U(1)$ current can be extracted from the expansion:
\eq{
\Sum_{\alpha\in \mathbb{Z}/l\mathbb{Z}}\psi^*_{\alpha}(z+t/2)\psi_{\alpha}(z-t/2)=\frac lt+J(z)+t T(z)+\ldots
}
Using \rf{definition1}, \rf{opehat} and \rf{definition2} one gets for the l.h.s.
\eq{
\Sum_{\alpha\in \mathbb{Z}/l\mathbb{Z}}\frac{\frac1l(z+\frac t2)^{\frac{1-l}{2l}} (z-\frac t2)^{\frac{1-l}{2l}}}{(z+\frac t2)^{1/l}-(z-\frac t2)^{1/l}}:e^{i\phi (e^{2\pi i\alpha}(z+t/2))-i\phi(e^{2\pi i\alpha}(z-t/2))}:\,=
\\
= \Sum_{\alpha\in \mathbb{Z}/l\mathbb{Z}}
\left(\frac 1t+t\frac{l^2-1}{24 l^2z^2}\right)e^{it\d\phi(e^{2\pi i\alpha} z)}+O(t^2)=\\=
\frac lt+\Sum_{\alpha\in \mathbb{Z}/l\mathbb{Z}}i\d\phi(e^{2\pi i\alpha} z)+\frac{t}{z^2}\frac{l^2-1}{24 l}-\frac t2\Sum_{\alpha\in \mathbb{Z}/l\mathbb{Z}}:\d\phi(e^{2\pi i\alpha} z)^2:+O(t^2)
}
One finds from here
\eq{
J(z)= \Sum_{\alpha\in \mathbb{Z}/l\mathbb{Z}}i\d\phi(e^{2\pi i\alpha} z) = \Sum_{k\in\mathbb Z}\frac{J_n}{z^{n+1}}
\\
T(z)=\frac{l^2-1}{24lz^2}+\frac1l\Sum_{k+n\in\mathbb Z}\frac{:J_nJ_k:}{z^{n+k+2}}
}
which already have expansions over the integer powers of $z$. Therefore
\eq{
L_0=\frac{l^2-1}{24l}+\frac1{2l}J_0^2+\frac1{l}\sum_{n>0}J_{-n}J_n \label{L0twisted}
}
and the character of this module is given by
\eq{
\tr q^{L_0}=q^{\frac{l^2-1}{24 l}}\frac{\Sum_{n\in\mathbb Z} q^{\frac{1}{2l}(lr+n)^2}}{\Prod_{n=1}^\infty(1-q^{\frac nl})}
\label{character1l}
}
Important detail here is the following: monodromy transformations of fermions over the whole cycle are given by
\eq{
\psi^*(e^{2\pi il}z)=(-1)^{l-1}e^{2\pi i J_0}\psi^*(z),\quad \psi(e^{2\pi il}z)=(-1)^{l-1}e^{-2\pi i J_0}\psi(z)
}
Here the factor $(-1)^{l-1}$ comes from the factors $z^{-\frac12}$ and $z^{\frac 1{2l}}$, whereas $e^{2\pi i J_0}$ comes from $z^{\frac1l J_0}$. Altogether
this means that twisted fermions above represent conjugacy class $[l,e^{2\pi i r}]_+$ with
\eq{
rl=J_0 \mod\mathbb Z\,.
}

\subsection{$l$ charged fermions --- standard bosonization}

From the modes \rf{modes} of the operators $\hat\psi(z)$, $\hat\psi^*(z)$ we can construct another $l$ fermions
\eq{
\psi_{(a)}(z)=\frac1{\sqrt{l}}\Sum_{p\in\mathbb Z+\frac12}\frac{\psi_{a+p}}{z^{a+p+\frac12}},\ \ \ \
\psi^*_{(a)}(z)=\frac1{\sqrt{l}}\Sum_{p\in\mathbb Z+\frac12}\frac{\psi_{-a+p}}{z^{-a+p+\frac12}}
}
where
\eq{
a\in\{\frac{l-1}{2l}+r,\frac{l-3}{2l}+r,\ldots,\frac{1-l}{2l}+r\}
}
These fermions can be bosonized in terms of $l$ ``normal'', untwisted, bosons
\eq{
\psi^*_{(a)}(z)=e^{i\varphi_{(a),-}(z)}e^{i\varphi_{(a),+}(z)}e^{Q_{(a)}}z^{J_{(a),0}}(-1)^{\Sum_{b<a}J_{(b),0}}\\
\psi_{(a)}(z)=e^{-i\varphi_{(a),-}(z)}e^{-i\varphi_{(a),+}(z)}e^{-Q_{(a)}}z^{-J_{(a),0}}(-1)^{\Sum_{b<a}J_{(b),0}}
\label{bosonization01}
}
where
\eq{
J_{(a),0}|0\rangle=a|0\rangle
}
Computation of the character in this case gives us
\eq{
\tr q^{L_0}=\frac{\Sum_{n_0,\ldots,n_{l-1}}{ q^{\Sum_{k=0}^{l-1}(r+\frac{1-l+2kl}{2l}+n_k)^2 }}}{\Prod_{n=1}^\infty(1-q^n)^l}\label{character2l}
}
One can easily see that equality between \rf{character1l} and \rf{character2l} follows from particular case of \rf{identity2}.

\newpage

\footnotesize

\bigskip
\noindent \textsc{Landau Institute for Theoretical Physics, Chernogolovka, Russia,\\
	Center for Advanced Studies, Skoltech, Moscow, Russia,\\
	Laboratory for Mathematical Physics, NRU HSE, Moscow, Russia,\\
	Institute for Information Transmission Problems, Moscow, Russia,\\
	Independent University of Moscow, Moscow, Russia}

\emph{E-mail}:\,\,\textbf{mbersht@gmail.com}\\

\noindent \textsc{Center for Advanced Studies, Skoltech, Moscow, Russia,\\
Laboratory for Mathematical Physics, NRU HSE, Moscow, Russia,\\
	Bogolyubov Institute for Theoretical Physics, Kyiv, Ukraine}

\emph{E-mail}:\,\,\textbf{pasha.145@gmail.com}\\

\noindent \textsc{Center for Advanced Studies, Skoltech, Moscow, Russia,\\
	Department of Mathematics and Laboratory for Mathematical Physics, NRU HSE, Moscow, Russia,\\
	Institute for Theoretical and Experimental Physics, Moscow, Russia\\
	Theory Department of Lebedev Physics Institute, Moscow, Russia}

\emph{E-mail}:\,\,\textbf{andrei.marshakov@gmail.com}

\end{document}